\documentclass[journal]{IEEEtran}

\usepackage{ucs}
\usepackage[utf8x]{inputenc}
\usepackage[cmex10]{amsmath}
\usepackage{cite, amsfonts, amssymb, bm, bbm, graphicx, relsize, multirow, booktabs, enumitem, algorithmic, algorithm, color, soul}
\usepackage[american]{babel}
\usepackage[T1]{fontenc}

\usepackage{amsthm}
\makeatletter 
\def\@endtheorem{\qed\endtrivlist\@endpefalse } 
\makeatother

\newtheorem{theorem}{Theorem}

\theoremstyle{definition}

\theoremstyle{remark}

\newcommand*{\herm}{^{\mathsf{H}}}
\newcommand*{\transp}{^{\mathsf{T}}}

\DeclareMathOperator*{\argmin}{\arg\min}

\newcommand{\e}{\mathrm{e}}
\renewcommand{\i}{\mathrm{i}}

\IEEEoverridecommandlockouts

\title{Semi-Blind Multi-Tag Ambient Backscatter Communications Using Radar Signals}

\author{
Luca~Venturino,~\IEEEmembership{Senior~Member,~IEEE}, Emanuele~Grossi,~\IEEEmembership{Senior~Member,~IEEE}, Jeremy~Johnston, Marco~Lops,~\IEEEmembership{Fellow,~IEEE},   Xiaodong~Wang,~\IEEEmembership{Fellow,~IEEE} 
\thanks{L.~Venturino and E.~Grossi are with the Department of Electrical and Information Engineering, University of Cassino and Southern Lazio, 03043 Cassino, Italy, and with CNIT, 43124 Parma, Italy (e-mail: l.venturino@unicas.it; e.grossi@unicas.it). M.~Lops is with the Department of Electrical and Information Technology, University of Naples Federico II, 80138 Naples, Italy, and with CNIT, 43124 Parma, Italy (e-mail: lops@unina.it). J.~Johnston and X.~Wang are with the Department of Electrical Engineering, Columbia University, New York, NY 10027, United States (e-mail: j.johnston@columbia.edu; xw2008@columbia.edu). }
\thanks{A preliminary version of this manuscript was presented at the 2023 IEEE International Workshop on Technologies for Defence and Security, Rome, Italy, and at the 2024 IEEE Radar Conference, Denver (CO), USA.}
\thanks{The work of M.~Lops was partially supported by the European Union under the Italian National Recovery and Resilience Plan (NRRP) of NextGenerationEU, partnership on ``Telecommunications of the Future'' (PE00000001 - program ``RESTART'' - E63C22002040007). The work of L.~Venturino and E.~Grossi was partially supported by the European Union -- NextGenerationEU -- National Recovery and Resilience Plan, Mission 4, Component 2, Investment 1.1, Call PRIN 2022 D.D. 104 02/02/2022 (Project 202238BJ2R ``CIRCE,'' CUP H53D23000420006).}
}

\begin{document}
\bstctlcite{BSTcontrol}	
\maketitle	\IEEEpeerreviewmaketitle

\begin{abstract}
In this work, we consider a backscatter communication system wherein multiple asynchronous sources (tags) exploit the reverberation generated by a nearby radar transmitter as an ambient carrier to deliver a message to a common destination (reader) through a number of available subchannels. We propose a new encoding strategy wherein each tag transmits both pilot and data symbols on each subchannel and repeats some of the data symbols on multiple subchannels. We then exploit this signal structure to derive two semi-blind iterative algorithms for joint estimation of the data symbols and the subchannel responses that are also able to handle some missing measurements. The proposed encoding/decoding strategies are scalable with the number of tags and their payload and can achieve different tradeoffs in terms of transmission and error rates. Some numerical examples are provided to illustrate the merits of the proposed solutions.
\end{abstract}

\section{Introduction}
Beyond 5G mobile communication networks are envisioned to provide massive connectivity and ubiquitous sensing capabilities, thus promoting the proliferation of Internet of Things (IoT) services; relevant applications include automotive, smart cities/factories, logistics, and warehousing~\cite{2021-6G-Machine, Nguyen-IoT-2022}. This technological leap requires enhancing the spectrum efficiency but also developing low-power and low-cost physical layer solutions to make the system sustainable in terms of consumed resources, required hardware, and electromagnetic pollution.

Techniques that address spectrum overcrowding range from spectrum sharing between two autonomous systems~\cite{Lops-2019}, to integrated sensing and communication (ISAC) architectures encompassing only one active transmitter and two different receiving chains to accommodate both functions~\cite{Hassanien-2016, Liu-2020}. On the other hand, ambient backscattering is a promising technology for implementing low-power and low-cost communications~\cite{Liu13ambientbackscatter}; indeed, a backscatter device (hereafter referred to as a \emph{tag}) can deliver a message to a receiver (hereafter referred to as a \emph{reader}) by reflecting and modulating an incident ambient signal emitted by the radio frequency (RF) emitter of an existing legacy system. Most studies have proposed to use the RF emitter of a nearby communication system, such as a TV tower, a Wi-Fi access point, and a cellular base station~\cite{Huynh2018, Zhang-2020}. To mitigate the direct interference from the RF source, several solutions have been proposed, including the design of suitable transceiver strategies and the exploitation of reconfigurable intelligent surfaces (RISs)~\cite{Tellambura-2016, Qian-2017a, Qian-2017b, Nawaz-2021, BackCom-RIS-2022}. 

Recent developments~\cite{2022-Venturino-Asilomar,2022-Venturino-ABC} have  shown that the probing signal of a nearby radar transmitter and the corresponding reverberation of the surrounding objects (refereed to as \emph{clutter} in keeping with the radar jargon) can also serve as an ambient carrier, providing  each tag with a set of communication subchannels sustained by the echoes of scattering centers with different bistatic ranges (i.e., with different radar-scatter-tag distances): interestingly, this implements a simplified form of radar-communication spectrum sharing. The key idea is that the locally periodic structure of the ambient carrier can be exploited for interference mitigation (whether from multiple tags or from the radar itself), even in the absence of channel state information (CSI). In particular, the tags can communicate with the reader on the same subchannel by resorting to a suitable intra-frame encoding/decoding rule or an inter-frame differential encoding/decoding rule, or both, with a frame being an interval spanning an integer number of radar periods and shorter than the coherence time of the propagation channel. The intra-frame codewords are forced to lie in the orthogonal complement of the subspace containing the radar interference: this has the merit of avoiding any signal loss at the price of limiting the maximum achievable transmission rate and of requiring a look-up encoding/decoding table. 

Considering the context outlined in~\cite{2022-Venturino-Asilomar,2022-Venturino-ABC} and starting from the preliminary results presented in~\cite{TechDefence2023, RadarConference2024}, the present contribution aims to introduce a novel intra-frame encoding/decoding strategy. At the transmit side, the tags undertake joint intra- and inter-subchannel encoding on a subset of subchannels at their disposal: intra-subchannel encoding consists of a concatenation of pilot and data symbols, while inter-subchannel encoding amounts to repeating some of the data symbols on multiple subchannels. The repeated data symbols can benefit from subchannel diversity and, therefore, experience better error protection. At the receive side, the reader exploits the underlying signal structure and resorts to semi-blind decoding procedures to iteratively estimate the data symbols and the subchannel responses; indeed, semi-blind methods can achieve significant performance gains with reduced training overhead as compared to a traditional signal recovery that first estimates the subchannel responses using only the pilot symbols and then uses the estimated subchannel responses to decode the data symbols~\cite{Carvalho2000,8807374, Wenjing-2019, Naraghi-Pour-2021}. A major advantage of the proposed encoding/decoding scheme over the one in~\cite{2022-Venturino-Asilomar,2022-Venturino-ABC} is that it is scalable with the number of tags and their payload, thus allowing a wider range of tradeoffs  between reliability, transmission rate, and implementation cost. More specifically, the main contributions of this study can be summarized as follows.
\begin{itemize}
\item After extending the signal model developed in~\cite{2022-Venturino-ABC} to the case of asynchronous tags and deriving a convenient matrix representation for the measurements collected by the reader over a given frame interval, we introduce in detail the aforementioned encoding scheme.

\item We provide sufficient conditions for noiseless recovery of the data symbols of all tags and their subchannel responses. These conditions set limits to the minimum number of pilot symbols assigned to each tag and reveal the existence of a tight interplay among the pilot symbols and the data symbols repeated across subchannels.

\item We derive two semi-blind decoding algorithms, which jointly process the measurement matrices at the output of the employed subchannels. Each algorithm iteratively solves a constrained regularized least squares problem to obtain a full rank factorization of each measurement matrix, wherein one factor contains the data symbols and the other one the subchannel response: the procedure monotonically converges to a stationary point. The novel feature here is that the involved constraints take advantage of the specific structure of the matrix factors induced by the proposed encoding rule and by the timing of the tags. Also, both solutions are robust against the radar and multi-tag interference and can handle some missing entries in the measurement matrices; their implementation cost is scalable with the number of tags and their payload.

\item We provide extensive analysis to verify the sufficient conditions for noiseless recovery, assess the impact of the main parameters on the bit error rate (BER) and the normalized root mean square error (NRMSE) in the estimation of the subchannel responses, and elicit the main tradeoffs among transmission and error rates, also in comparison with the encoding/decoding strategy in~\cite{2022-Venturino-ABC},
the semi-blind decoding method in \cite{8807374}, and a genie-aided maximum likelihood (ML) decoder with a perfect channel state information (CSI).
\end{itemize}

The remainder of the paper is organized as follows. Sec.~\ref{SEC:System-description} contains the system description and the signal model. Sec.~\ref{SEC:encoding} presents the proposed encoding scheme. Sec.~\ref{SEC:algorithms} illustrates the proposed decoding algorithms. Sec.~\ref{SEC:Numerical analysis} contains some numerical results. Sec.~\ref{SEC:Conclusions} gives the concluding remarks. Finally, we defer to the Appendix for some mathematical derivations.

\paragraph*{Notation}  In the following, $\mathbb C$ is the set of complex numbers. Column vectors and matrices are denoted by lowercase and uppercase boldface letters, respectively. The symbols $\Re\{\,\cdot\,\}$,   $(\,\cdot\,)\transp$, and $(\,\cdot\,)\herm$ denote real part, transpose, and conjugate-transpose, respectively. $\bm{1}_{M}$ and $\bm{0}_{M}$, are the $M$-dimensional all-one and all-zero column vectors, respectively.   $\bm{I}_{M}$ is the $M\times M$ identity matrix.   $\bm{O}_{M,N}$ is the $M\times N$ all-zero matrix.  $\|\bm{A}\|_F$, $\mathrm{rank}\{\bm{A}\}$, $\mathrm{col}\{\bm{A}\}$, and $\mathrm{null}\{\bm{A}\}$ are the Frobenius norm, the rank, the column space, and the null space of the matrix $\bm{A}$. $\mathrm{vec}\{\bm{A}\}$ is the vector obtained by stacking up the columns of the matrix $\bm{A}$. $[\bm{A}]_{i,j}$ denotes the entry in the $i$-th row and $j$-th column of the matrix $\bm{A}$. $[\bm{A}]_{:,a:b}$ and $[\bm{A}]_{a:b,:}$ denote the submatrices of $\bm{A}$ containing only the columns with indexes from $a$ to $b$ and the rows with indexes from $a$ to $b$, respectively. $[\bm{A}_{1} \ \cdots \ \bm{A}_{n}]$ and $[\bm{A}_{1}; \ \cdots \ ;  \bm{A}_{n}]$ denote the horizontal and vertical concatenations of the matrices $\bm{A}_{1},\ldots,\bm{A}_{n}$. $\mathrm{diag}(\bm{a})$ and $\mathrm{diag}\bigl(\{a_n\}_{n=1}^{N}\bigr)$ denote the diagonal matrix with the entries of the vector $\bm{a}=(a_1 \ \cdots \ a_{N})\transp$ on the main diagonal. For sets $A$ and $B$, $A+B$ is the Minkowski sum, i.e., $A+B= \{a+b \ | \ a\in A, b\in B\}$. Finally, $\i$, $\star$, $\otimes$, $\odot$, and $\mathrm{E}[\,\cdot\,]$ denotes the imaginary unit, the convolution operator, the Kronecker product, the Schur product, and the statistical expectation, respectively. 

\section{System Description}\label{SEC:System-description}
\begin{figure}[!t]
\centerline{\includegraphics[width=\columnwidth]{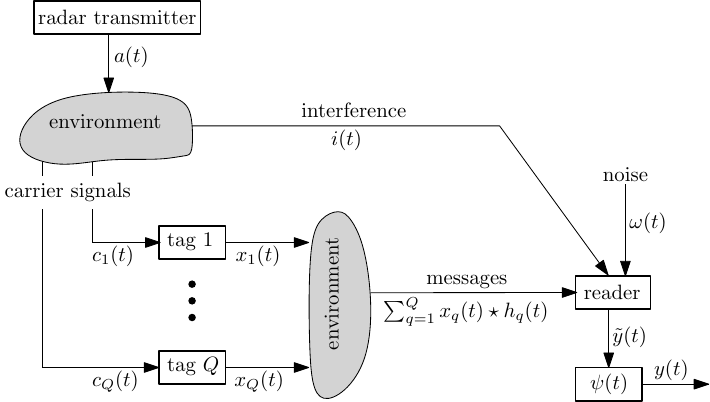}}
\caption{Graphical illustration of the considered radar-enabled ambient backscatter communication system.}
	\label{fig_1}
\end{figure}

Consider a system encompassing a radar transmitter, $Q$ asynchronous single-antenna tags, and a single-antenna reader, as outlined in Fig.~\ref{fig_1}. The radar emits the signal $\Re\{a(t)\e^{\i 2\pi f_{a}t}\}$, where $f_{a}$ is the carrier frequency and  $a(t)$ is a baseband periodic waveform of period $T_{a}$ and bandwidth $W_{a}$; this excitation propagates through a rich scattering environment, and the resulting echoes (possibly including the direct signal from the radar transmitter) reach both the tags and the reader. The tags employ the incident clutter as a carrier signal to send a message to the reader. In the following, the knowledge of $a(t)$ is not required, and the tags and reader only know its period. Message encoding/decoding is carried out over a time segment spanning $L$ consecutive radar periods, which is referred to as a \emph{frame interval};  without loss of generality, the frame interval $[0, L T_{a})$ is considered for illustration. Finally, we assume that the responses of the radar-tag, radar-reader, and tag-reader propagation channels remain constant in each frame interval, whereby they can be modeled as causal linear time-invariant filters in what follows.

\subsection{Signal backscattered by the tag}
Let $\Re\{c_q(t)\e^{\i 2\pi f_{a}t}\}$ be the continuous-time radar clutter hitting the $q$-th tag over the frame interval $[0,L T_{a})$, for $q=1,\ldots,Q$, where $c_q(t)$ is its baseband representation. Under the considered assumptions, we have
\begin{equation}\label{tag-clutter}
	c_q(t)=\int_{0}^{\infty}\gamma_q(\tau) a(t-\tau)d\tau,
\end{equation}
for $t\in[0,L T_{a})$, where $\gamma_q(t)$ is the baseband impulse response of the radar-tag channel, whereby $c_q(t)$ presents $L$ equal cycles of duration $T_a$. The backscatter modulator alters the incident waveform by changing its phase and/or amplitude. The baseband representation of the continuous-time signal backscattered by the $q$-th tag over the frame interval is~\cite{2022-Venturino-ABC}
\begin{align}
	x_{q}(t)&=\sum_{\ell=1}^{L} \sum_{n=1}^{N_{s}}x_{q,\ell,n} \notag \\   &\quad \times\underbrace{c_{q}(t) \Pi\left(\frac{t-(\ell-1) N_{s}T_s+(n-1)T_s}{\Delta_{s}}\right)}_{\phi_{q,\ell,n}(t)},\label{tag-signal}
\end{align}
for $t\in[0,L T_{a})$, where $N_s$ is the number of symbols sent in one radar period, $T_{s}=T_a/N_{s}$ is the symbol interval, $x_{q,\ell,n}$ is the symbol sent in $n$-th symbol interval of the $\ell$-th radar period, which belongs to a given alphabet $\mathcal{X}$, $\Pi(t)$ is a rectangular pulse which has unit amplitude  for $t\in[0,1)$ and is zero elsewhere, and $\Delta_{s}=T_{s}-\Delta_g$ is the symbol duration, with $\Delta_g$ being a guard interval between the transmission of consecutive symbols. Fig.~\ref{fig_2}a provides a graphical description of $x_{q}(t)$ in~\eqref{tag-signal}; we underline that this signal only accounts for the antenna mode scattering of the tag, which is varied by acting on the impedance of the antenna load~\cite{BackCom-RIS-2022}.

\subsection{Signal received by the reader}
The baseband continuous-time signal received by the reader over the frame interval is
\begin{equation}\label{reader-signal}
\tilde{y}(t) = \sum_{q=1}^{Q}x_q(t) \star h_q(t) + i(t) + \omega(t),
\end{equation}
for $t\in[0, L T_{a})$, where $h_q(t)$ is the baseband continuous-time impulse response of the channel between the $q$-th tag and the reader (also accounting for the tag scattering efficiency and any carrier phase offset), which is zero outside $[\tau_{h,q}, \tau_{h,q}+\Delta_{h})$, $i(t)$ is the interference produced by the radar transmitter (including the structural mode scattering of the tags),  and $\omega(t)$ is the additive noise. The delay spread $\Delta_h$  of the tag-reader channels is known and assumed to be the same for all channels, while the delays $\tau_{h,1},\ldots,\tau_{h, Q}$ account for some timing offset among the tags and are unknown. The radar clutter $i(t)$ can be modeled as in \eqref{tag-clutter} and again presents $L$ equal cycles of duration $T_a$ over the processed interval. 

\begin{figure*}[!t]
\centerline{\includegraphics[width=0.9\textwidth]{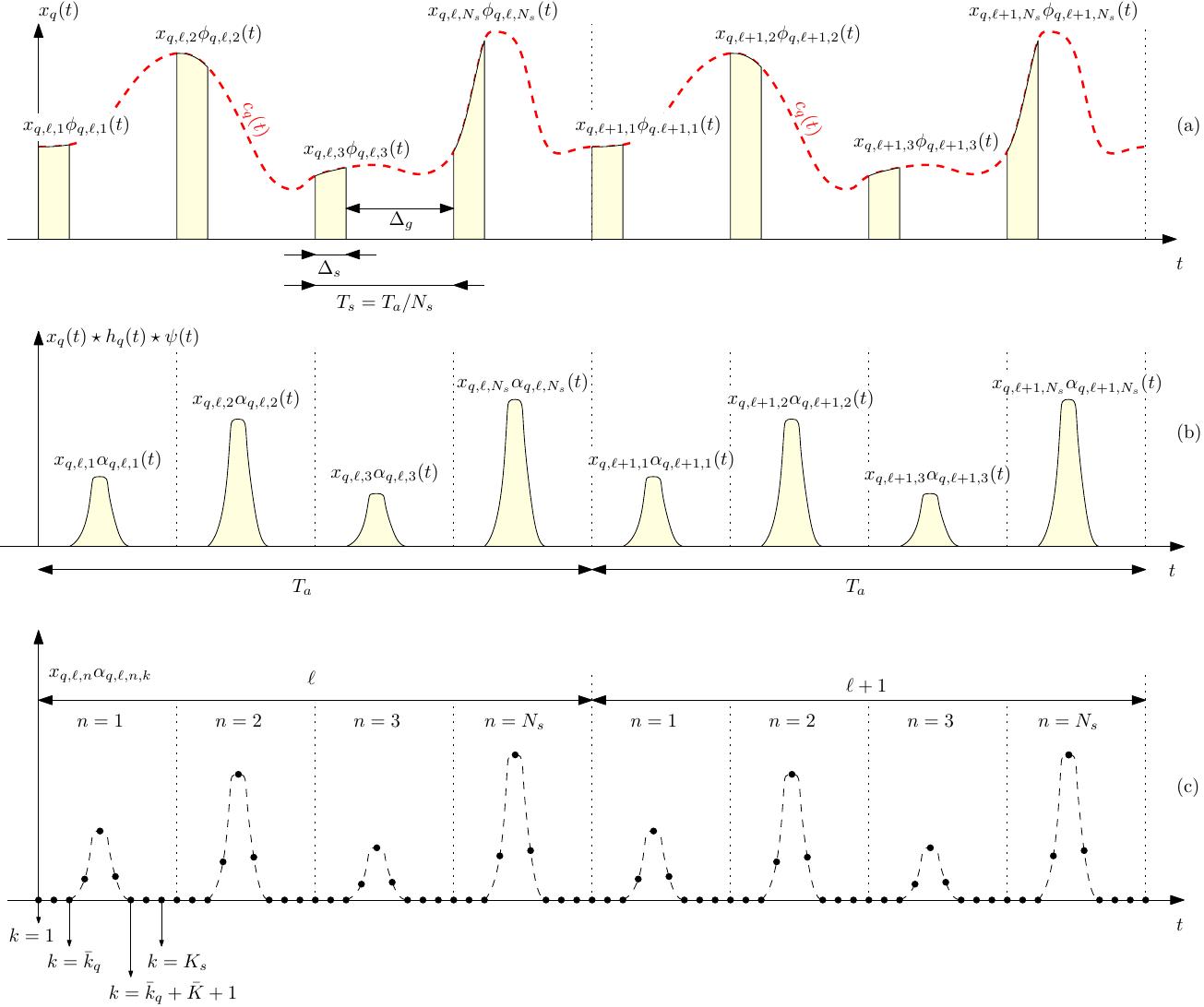}}
	\caption{Example of the signal $x_{q}(t)$ backscattered by the $q$-th tag (a), the received waveform $x_q(t)\star h_q(t) \star \psi (t)$ at the output of the receive filter over two radar periods (b), and of the corresponding samples (c) when $L\geq2$, $N_s=4$, $\bar{k}_q=2$, $\bar{K}=3$, and $K_s=9$.}
	\label{fig_2}
\end{figure*}
The signal $\tilde{y}(t)$ is passed through a low-pass filter, whose impulse response $\psi(t)$ is zero outside $[0,\Delta_\psi)$ and has unit energy and bandwidth $W_\psi$; the corresponding output is 
\begin{align}
	y(t)&=\sum_{q=1}^{Q}\sum_{\ell=1}^{L} \sum_{n=1}^{N_{s}}x_{q,\ell,n}   \alpha_{q,\ell,n}(t)+ i(t)\star \psi (t) \notag \\ &\quad  + \omega(t) \star \psi (t), \label{rx_signal_1}
\end{align}
for $t\in[0,L T_{a})$, where $\alpha_{q,\ell,n}(t)= \phi_{q,\ell,n}(t) \star h_{q}(t) \star \psi(t)$ is the unknown received pulse carrying the symbol $x_{q,\ell,n}$. In order to avoid intersymbol interference, the guard interval is chosen to satisfy $ \Delta_g \geq \tau_h + \Delta_h+ \Delta_\psi$,  where $\tau_h$ is a known upper bound to the delay offsets of the tags.  Fig.~\ref{fig_2}b illustrates the waveform $ x_q(t)\star h_q(t) \star \psi (t)$: it is seen that the modulated pulses $\{x_{q,\ell,n}\alpha_{q,\ell,n}(t)\}$ do not overlap in time. 

The filtered signal $y(t)$ is sampled at rate $K_s/T_s$ over $[0,L T_{a})$, producing the samples
\begin{align}
	y_{\ell,n,k} &= y(t)\big|_{t=(\ell-1)N_{s}T_{s}+(n-1)T_{s}+(k-1) T_s/K_s} \notag \\
	&= \sum_{q=1}^{Q} x_{q,\ell,n} \underbrace{\alpha_{q,\ell,n}(t)\big|_{t=(\ell-1)N_{s}T_{s}+(n-1)T_{s}+(k-1) T_s/K_s}}_{\alpha_{q,\ell,n,k}} \notag\\
	& \quad + \underbrace{i(t) \star \psi(t)\big|_{t=(\ell-1)N_{s}T_{s}+(n-1)T_{s}+(k-1) T_s/K_s}}_{i_{\ell,n,k}} \notag\\
	& \quad + \underbrace{ \omega(t)\star\psi(t)\big|_{t=(\ell-1)N_{s}T_{s}+(n-1)T_{s}+(k-1) T_s/K_s}}_{ \omega_{\ell,n,k}},\label{rx_signal_2}
\end{align}
for $\ell=1,\ldots,L$, $n=1,\ldots,N_s$, and $k=1,\ldots,K_s$,  where $K_s  \geq \lceil W_{\psi}T_{s}\rceil$. The periodicity of the signals $c_q(t)$ and $i(t)$ over a frame interval implies that 
\begin{subequations}
	\begin{align}
		\alpha_{q,1,n,k}&=\cdots=\alpha_{q,L,n,k},\\
		i_{1,n,k}&=\cdots=i_{L,n,k}.
	\end{align}
\end{subequations}
To take advantage of this signal structure, the above samples are parsed into $N_s$ groups, which define as many time-orthogonal subchannels;  the $n$-th subchannel contains the observations taken in the $n$-th symbol interval of each radar period, which are organized into the following matrix
\begin{align}
	\bm Y_{n}&=	\begin{bmatrix}y_{1,n,1 }&\ldots &  y_{1,n,K_s} \\
		\vdots & & \vdots \\
		y_{L,n,1} &\ldots & y_{ L,n,K_s}
	\end{bmatrix}\notag\\&=\sum_{q=1}^{Q}\bm{x}_{q,n}\bm{\alpha}_{q,n}\transp+\bm{1}_{L}\bm{i}_{n}\transp+\bm{\Omega}_{n} \in \mathbb{C}^{L\times K_s}, \label{rx_signal}
\end{align}
for $n=1,\ldots,N_s$, where 
\begin{subequations}
	\begin{align}		\bm{x}_{q,n}&=\begin{bmatrix}x_{q,1,n}&\cdots&x_{q,L,n}\end{bmatrix}\transp\in\mathcal{X}^{L},\\
		\bm{\alpha}_{q,n}&=\begin{bmatrix}\alpha_{q,1,n,1}&\cdots&\alpha_{q,1,n,K_s}\end{bmatrix}\transp \in\mathbb{C}^{K_s},\\
		\bm{i}_n&=\begin{bmatrix}i_{1,n,1}&\cdots&i_{1,n,K_s}\end{bmatrix}\transp \in\mathbb{C}^{K_s},
	\end{align}
\end{subequations}
while $\bm{\Omega}_{n}$ is defined similarly to $\bm Y_{n}$. 

The following remarks are now in order. The vector $\bm{x}_{q,n}$ is the codeword sent by the $q$-th tag over $L$ consecutive uses of the $n$-th subchannel, which may contain both pilot and data symbols (more on this in Sec.~\ref{SEC:encoding}), while the vector $\bm{\alpha}_{q,n}$ can be regarded as the corresponding subchannel response. Some entries of $\bm{\alpha}_{q,n}$ are zeros, as a consequence of the gating operation induced by the antenna mode scattering of the tag; indeed, as illustrated in Fig.~\ref{fig_2}c,  $\alpha_{q,\ell,n,k}=  0$ if   $k=1, \ldots, \bar{k}_q$  or $k=\bar{k}_q+\bar{K}+1,\ldots, K_s$, where $\bar{k}_q$ and $\bar{K}$ are tied to $\tau_{h,q}$ and  $\Delta_{s}+\Delta_{h}+\Delta_{\psi}$, respectively, with  $\bar{k}_q\geq 0$, $\bar{K}\geq 1$, and $\bar{k}_q+\bar{K}\leq K_{s}$. Upon exploiting this property, we can write
\begin{equation}\label{A-structure}
\bm{\alpha}_{q,n}=\bm{S}(\bar{k}_q)\bar{\bm{\alpha}}_{q,n},
\end{equation}
where $\bm{S}(k)=[\bm{O}_{k, \bar{K}}; \bm{I}_{\bar{K}};\bm{O}_{K_s-\bar{K}-k, \bar{K}}]$ is a $K_s\times \bar{K}$ shift matrix and 
\begin{equation}	\bar{\bm{\alpha}}_{q,n}=\begin{bmatrix}\alpha_{q,1,n,\bar{k}_q+1}&\cdots&\alpha_{q,1,n,\bar{k}_q+\bar{K}}\end{bmatrix}\transp \in\mathbb{C}^{\bar{K}}.
\end{equation}
Finally, the radar interference $\bm{1}_{L}\bm{i}_{n}\transp$ can be regarded as the (unmodulated) signal backscattered by a virtual tag,  say tag $Q+1$, that always sends the all-one codeword, with $\bm{i}_{n}$ being the corresponding subchannel response; unlike $\bm{\alpha}_{q,n}$, there is no specific structure in $\bm{i}_{n}$.

The receiver signal in~\eqref{rx_signal} can also be recast as 
\begin{equation}\label{rx_signal_Yn}
	\bm Y_{n}=\bm{X}_{n}\bm{A}_{n}  +\bm{\Omega}_{n} \in\mathbb{C}^{L\times K_s},
\end{equation}
where $\bm{X}_{n}=\bigl[\bm{x}_{1,n} \ \cdots \ \bm{x}_{Q,n} \ \bm{1}_{L}\bigr]$ and $\bm{A}_{n}=\bigl[\bm{\alpha}_{1,n} \ \cdots \ \bm{\alpha}_{Q,n} \ \bm{i}_{n}\bigr]\transp$, for $n=1,\ldots,N_s$. Hereafter, we refer to $\bm{Y}_{n}$, $\bm{X}_{n}$, $\bm{A}_{n}$, and $\bm{\Omega}_{n}$ as the measurement matrix, the symbol matrix, the subchannel response matrix, and the noise matrix, respectively. Also, we assume $\mathrm{rank}\big(\bm{X}_{n}\big)=\mathrm{rank}\big(\bm{A}_{n}\big)=Q+1\leq \min\{L, K_s\}$, which is a necessary condition for noiseless data recovery. Notice that $\mathrm{rank}\big(\bm{X}_{n}\big)=Q+1$ can be ensured through a suitable encoding strategy (see Sec.~\ref{SEC:encoding}), while the condition on $\bm A_n$ requires more discussion. If the active tags have different delays (i.e., $\bar{k}_i\neq \bar{k}_j$ for any $i\neq j$), then $\mathrm{rank}\big(\bm{A}_{n}\big)=Q+1$ always holds provided that $K_s \geq Q+1$. On the other hand, determining the conditions for $\bm A_n$ to be full-rank in the presence of synchronous tags requires some understanding of the physics of the scattering process. Assume for simplicity that the subchannel responses $\{h_q(t)\}_{q=1}^Q$ are impulsive, that the receive filter $\psi(t)$ is an ideal low-pass one, and that the sampling rate is equal to the radar bandwidth. The clutter signal from a single bistatic range cell\footnote{A bistatic range cell is the set of points with unresolvable propagation delays as measured from the radar transmitter to the tag.} is thus an undistorted replica of the radar probing signal, weighted by a random complex gain encapsulating the global scattering characteristics pertaining to that cell. The use of a symbol duration spanning $B_a$ range resolution cells (i.e., $\Delta_s=B_a/W_a$) results in a clutter-generated signature which is a linear combination of $B_a$ linearly independent vectors, each weighted by the complex gain of the corresponding bistatic range cell. Thus, a set of $Q$ vectors having this structure can be linearly independent only if $B_a \geq Q \geq \bar K$. This line of thought carries over to the case that $\{h_q(t)\}_{q=1}^Q$ are not impulsive and $\psi(t)$ is a non-ideal low-pass filter.

\section{Encoding Scheme} \label{SEC:encoding}
We perform encoding across a subset of ${N}$ subchannels, with $N\in\{1,\ldots,N_s\}$.  To simplify exposition, the subchannels indexed by $1,\ldots, {N}$ are employed. We propose the following encoding rule
\begin{equation}
	{\bm X}_{n} = \begin{bmatrix}
		{\bm P}_{n} & {\bm 1}_{P_n} \\  
            \bm D_{0} & {\bm 1}_{D_0} \\  
            \bm D_{n} & {\bm 1}_{D_n}
		\end{bmatrix},
  \label{eq:Xnstructure}
\end{equation}
where ${\bm P}_{n}\in \mathcal{X}^{P_{n}\times Q}$ is the matrix containing the pilot symbols sent by the tags on the $n$-th subchannel, ${\bm D}_{0}\in \mathcal{X}^{D_{0}\times Q}$ is the matrix containing the data symbols repeated by the tags on all subchannels, and ${\bm D}_{n}\in \mathcal{X}^{D_{n}\times Q}$ is the matrix containing the data symbols sent by the tags only on the $n$-th subchannel, with $P_n+D_0+D_n=L$, and $n=1,\ldots, N$. Notice that $P=\sum_{n=1}^{N}P_{n}$ and $D=\sum_{n=0}^{N}D_{n}$  are the number of pilot and data symbols sent by each tag in $L$ uses of the $N$ subchannels, respectively; also, for future reference, we define the pilot symbol matrix  $\bm P=[\bm P_{1}; \cdots; \bm P_{N}]\in\mathcal{X}^{P\times Q}$ and the data symbol matrix $\bm D=[\bm D_0;	\bm D_1;\cdots;	\bm D_{N}]\in\mathcal{X}^{D\times Q}$.

For the considered encoding rule, the transmission rate is\footnote{Since each subchannel is used only once in each radar period, \eqref{eq:user-rate} is also the number of bits per radar period and per subchannel sent by each tag.}
\begin{equation}
R=\frac{D}{NL}\log_2|\mathcal X|,\quad \text{[bits/subchannel-use/tag]}.\label{eq:user-rate}
\end{equation}
For fixed $L$ and $N$, different tradeoffs in terms of transmission and error rates can be obtained by varying $P$, $D_0$, and $\mathcal X$ (more on this in Sec.~\ref{SEC:Numerical analysis}). When $D_0=D$, each tag sends the same data symbols in all subchannels, thus taking full advantage of the subchannel diversity (if any). On the other hand, when $D_0=0$, independent data symbols are sent across the subchannels, thus maximizing the transmission rate: in this case, we have a set $N$ of independent subchannels, and there is no benefit in jointly processing the corresponding outputs. For the same transmission rate, $P$ needs to be reduced if $D_0$ is increased or $|\mathcal X|$ is reduced.

\subsection{Noiseless unique data recovery}
 We assume in this subsection that no noise is present in the measurement matrix $\bm Y_n$ and $\mathrm{rank}(\bm Y_n)=Q+1$, for $n=1,\ldots,N$. Suppose the reader recovers a full rank factorization~\cite{piziak1999} of $\bm Y_n$, namely full rank matrices $\bm T_n$ and $\bm V_n$ with sizes $L\times (Q+1)$ and $(Q+1)\times K_s$, respectively, such that $\bm Y_n= \bm T_n \bm V_n$, for $n=1,\ldots,N$.
The question is whether the full rank factorizations $\{\bm T_n \bm V_n\}_{n=1}^{N}$ are unique when the left factors possess the structure in~\eqref{eq:Xnstructure}, i.e., 
\begin{equation}
	{\bm T}_{n} = \begin{bmatrix}
		{\bm P}_{n} & {\bm 1}_{P_n} \\  
            \bm U_{0} & {\bm 1}_{D_0} \\  
            \bm U_{n} & {\bm 1}_{D_n}
		\end{bmatrix}
  \label{eq:Tnstructure}
\end{equation}
for some $\bm{U}_0\in\mathcal{X}^{D_0\times Q}$ and $\bm{U}_n\in\mathcal{X}^{D_n\times Q}$, while the first $Q$ rows of the right factors have the structure in~\eqref{A-structure}, i.e., 
\begin{equation}
\bm{V}_n=\left[\bm{S}(\bar{d}_{1})\bar{\bm{v}}_{1,n} \ \cdots \ \bm{S}(\bar{d}_{Q}) \bar{\bm{v}}_{Q,n} \ \bm{g}_{n}\right]\transp
\end{equation}
for some $\bar{d}_{q}\in\{0,\ldots,\lceil\tau_{h} K_s/T_s\rceil\}$, $\bar{\bm{v}}_{q,n}\in \mathbb{C}^{\bar K}$ and $\bm{g}_n\in \mathbb{C}^{K_s}$. If they are unique, we can conclude that $\bm T_n = \bm X_n$ and $\bm V_n = \bm A_n$, for $n=1,\ldots,N$, and, therefore, $\bm{U}=[\bm U_0;	\bm U_1;\cdots;	\bm U_{N}]=\bm{D}$.

Finding necessary and sufficient conditions for noiseless, unique data recovery is a challenging problem. Upon relying on the signal structure induced by the pilot symbols in $\bm{P}$ and by the repeated data symbols in $\bm{D}_{0}$, the following theorem provides sufficient conditions; the proof is in the Appendix.
\begin{theorem}
\label{thm:unique}
Suppose that the noiseless measurement matrix $\bm Y_n$ has rank $Q+1$, and that $\bm T_n \bm V_n$ is a given full rank factorization of $\bm Y_n$ with ${\bm T}_{n}$ as in~\eqref{eq:Tnstructure}, for $n=1,\ldots,N$. Then, these factorizations are unique if
\begin{subequations}\label{thconds}
\begin{gather}   
    \mathrm{rank}\bigl( [ \bm P \;  \bm 1_{P} ] \bigr)= Q+1, \label{thconds-1}\\
     \exists m:    \Bigl(\mathrm{null} \bigl( [ \bm P_n \; \bm 1_{P_n}]\bigr) + \mathrm{null} \bigl( [ \bm P_m \; \bm 1_{P_m}]\bigr)\Bigr) \notag \\ \cap \, \mathrm{null} \bigl( [ \bm D_0\; \bm 1_{D_0}]\bigr)=\{\bm 0_{Q+1}\}, \quad \forall n. \label{thconds-2}  
    \end{gather}
\end{subequations}%
\end{theorem}

\subsection{Some insights on Theorem~\ref{thm:unique} }
Notice first that the columns of the matrix $[ \bm P \;  \bm 1_{P} ]$ contain the pilot sequences assigned to each tag (including the all-one vector of the virtual tag); condition~\eqref{thconds-1} is tantamount to requiring that these pilot sequences are linearly independent; hence, we must necessarily have $P\geq Q+1$. 

Next, observe that each tag partitions its pilot sequence of length $P$ into $N$ pilot subsequences of length $P_1,\ldots,P_N$, each sent on a different subchannel. Condition~\eqref{thconds-2} establishes a connection between the structure of the matrix $[\bm P_n \; \bm 1_{P_n}]$ containing the $Q+1$ pilot subsequences sent on subchannel $n$, for $n=1,\ldots, N$, and the structure of the matrix $[ \bm{D}_0 \;  \bm 1_{D_0} ]$  containing the data symbols repeated across all subchannels. In particular, we have that 
\begin{align}
\eqref{thconds-2} & \Longrightarrow  \mathrm{null} \bigl( [\bm D_0\; \bm 1_{D_0}]\bigr)\cap\mathrm{null} \bigl( [\bm P_n \; \bm 1_{P_n}]\bigr)=\{\bm 0_{Q+1}\}, \;  \forall n \notag \\
& \Longleftrightarrow \mathrm{null}\bigl( [\bm P_n \; \bm 1_{P_n} ; \bm D_0\; \bm 1_{D_0}]\bigr)=\{\bm 0_{Q+1}\}, \;  \forall n \notag\\
&\Longleftrightarrow \mathrm{rank} \bigl( [\bm P_n \; \bm 1_{P_n} ; \bm D_0\; \bm 1_{D_0}]\bigr)=Q+1, \;  \forall n.\label{eq:thconds-3}
\end{align}
Hence,~\eqref{thconds-2} implies that the matrix $[\bm P_n \; \bm 1_{P_n} ; \bm D_0\; \bm 1_{D_0}]$, obtained by the vertical concatenation of the pilot and repeated data symbols on subchannel $n$, is full rank, for $n=1,\ldots, N$. 

If $\mathrm{rank}([\bm P_n \ \bm 1_{P_n}]) = Q+1$ for $n=1,\ldots,N$, the conditions in~\eqref{thconds} are satisfied regardless of $\bm D_0$. Interestingly, the tags may reuse here the same pilot subsequences on multiple subchannels. A limitation of this encoding scheme is that each tag must send at least $(Q+1)$ pilot symbols per subchannel. When $D_0=0$, this is the only way to meet the conditions in~\eqref{thconds}. When $D_0\geq1$, sending less than $(Q+1)$ pilot symbols per subchannel is possible, as highlighted next.

Assume that condition~\eqref{thconds-1} is true and that there is at least one subchannel without pilot symbols. In this case, condition~\eqref{thconds-2} is satisfied if $\mathrm{rank}([\bm D_0 \ \bm 1_{D_0}]) = Q+1$. To get more insight on this scenario, consider $N=2$, $\mathrm{rank}([\bm P_1 \ \bm 1_{P_1}]) = \mathrm{rank}([\bm D_0 \ \bm 1_{D_0}])= Q+1$, and $P_2=0$: the intuition is that the pilot subsequences on subchannel~$1$ can be used to estimate the unknown subchannel response matrix $\bm{A}_{1}$ and recover $\bm{D}_{0}$ and $\bm{D}_{1}$; then, the estimated data symbols in $\bm{D}_{0}$ can work as pilot symbols on subchannel~$2$ and, therefore, can be used to estimate the unknown subchannel response matrix $\bm{A}_{2}$ and recover $\bm{D}_{2}$. A key point is that, even if  $D_{0}\geq Q+1$, the data-dependent matrix $[\bm D_0 \ \bm 1_{D_0}]$ can be rank deficient in some frames when $\mathcal{X}$ is discrete, and the entries of $\bm D_0$ are independently drawn from $\mathcal{X}$; in practice, this occurrence can be neglected only when it is less frequent than other type of errors, as for example those due to the additive noise, the multi-tag interference, and the imperfect estimation of the subchannel matrices (more on this in Sec.~\ref{SEC:Numerical analysis-multichannel} when discussing Fig.~\ref{BER_NRMSE_vs_D0}). 

Finally, notice that the conditions in~\eqref{thconds} are satisfied if 
\begin{subequations}\label{thconds-rank}
\begin{align}   
   &\exists m:  \mathrm{rank}\bigl( [ \bm P_m \;  \bm 1_{P} ] \bigr)= Q+1, \label{thconds-rank-1}\\
    & \mathrm{rank} \bigl( [\bm P_n \; \bm 1_{P_n} ; \bm D_0\; \bm 1_{D_0}]\bigr)=Q+1, \;  \forall n. \label{thconds-rank-2} 
    \end{align}
\end{subequations}%
Indeed, condition~\eqref{thconds-1} is implied by~\eqref{thconds-rank-1}; also, since $\mathrm{null}([\bm P_m \ \bm 1_{P_m}])=\{\bm 0_{Q+1}\}$, condition~\eqref{thconds-2} becomes 
\begin{equation}
\mathrm{null} \bigl( [ \bm D_0\; \bm 1_{D_0}]\bigr) \cap \mathrm{null} \bigl( [ \bm P_n \; \bm 1_{P_n}]\bigr)=\{\bm 0_{Q+1}\}, \quad \forall n
\end{equation}
that is implied by~\eqref{eq:thconds-3}. We underline that the conditions in~\eqref{thconds-rank} are more restrictive than those in~\eqref{thconds}, but, in some cases, they may be checked more easily. For example, if $P_1=3$, $P_2=2$, $D_0=3$, and $Q=3$, then the following matrices (whose rows are conveniently taken from the $4\times 4$ Hadamard matrix) satisfy~\eqref{thconds} but not~\eqref{thconds-rank}:
\begin{subequations}
\begin{align}
[\bm{P}_1 \ \bm{1}_{P_1}]&=
\begin{bmatrix}
  1 & 1 & 1 & 1\\
 -1 &-1 & 1 & 1\\
 -1 & 1 &-1 & 1
\end{bmatrix},\\
[\bm{P}_2 \ \bm{1}_{P_2}]&=
\begin{bmatrix}
 1 &-1 &-1 & 1\\
 1 & 1 & 1 & 1
\end{bmatrix},\\
[\bm{D}_0 \ \bm{1}_{D_0}]&=
\begin{bmatrix}
 -1 &-1 & 1 & 1\\
 -1 & 1 &-1 & 1\\
  1 &-1 &-1 &1
\end{bmatrix}.
\end{align}
\end{subequations}

\section{Decoding Algorithms}\label{SEC:algorithms}
We propose here two semi-blind iterative algorithms to recover the data symbol matrix $\bm{D}$ and the subchannel response matrices $\{{\bm A}_{n}\}_{n=1}^{N}$, wherein the former only exploits the prior knowledge on structure of the symbol matrices, while the latter also takes into account the prior knowledge on the structure of the subchannel matrices. Interestingly, both algorithms can be  adapted to handle missing entries in the measurement matrices. Before proceeding, notice that~\eqref{eq:Xnstructure} can also be written as
\begin{equation}
	{\bm X}_{n} = \begin{bmatrix}
		{\bm P}_{n} & {\bm 1}_{P_n} \\  \bm F_n \bm D & {\bm 1}_{D_0+D_n}
		\end{bmatrix},
  \label{eq:Xnstructure-1}
\end{equation}
where $\bm F_n \in \{0,1\}^{(D_0+D_n) \times D}$ is a selection matrix such that $\bm F_n \bm D = [\bm D_0; \bm D_n]$: we will rely on the above more compact expression in the derivations of the proposed algorithms.  

\subsection{Algorithm 1} \label{alg_1_sec}
We propose here to solve the following constrained regularized least squares problem~\cite{regularized_ALS}
\begin{subequations} \label{P1-multi}
	\begin{align}
		\underset{\substack{{\bm U}\in\mathcal{X}^{D\times Q}\\ \bm V_n \in \mathbb{C}^{(Q+1)\times K_s}\,\forall n}}{\min}&\sum_{n=1}^{N}\|\bm Y_n - \bm T_n \bm V_n\|_F^2 \notag \\[-10pt] &\quad + \lambda_{u}\|\bm U\|_F^2 +  \sum_{n=1}^{N} \lambda_{v,n}\|\bm V_n\|_F^2,  \\[10pt]
		\mathrm{s.t.} \; 
        &\bm T_n=
		\begin{bmatrix}
			{\bm P}_{n} & {\bm 1}_{P_n} \\ 
			{\bm F}_{n} {\bm U} & {\bm 1}_{D_0+D_n}
		\end{bmatrix}, \; \forall n,\label{P1-multi-C2}
 \end{align}%
\end{subequations}%
where $\lambda_{u}\geq 0$ and $\lambda_{v,n}\geq 0$ are the regularization parameters. Notice that the variables ${\bm U}$, ${\bm T}_{n}$ and ${\bm V}_{n}$ represent an estimate of the true data symbol matrix ${\bm D}$,  of the true symbol matrix ${\bm X}_{n}$, and of the true subchannel response matrix ${\bm A}_{n}$, respectively, for $n=1,\ldots,N$. According to the underlying encoding rule, the constraint in~\eqref{P1-multi-C2} ensures that ${\bm T}_{n}$ is uniquely specified by ${\bm U}$, for $n=1,\ldots,N$; meanwhile, no structure for ${\bm V}_{n}$ is explicitly enforced. 

A suboptimal solution to Problem~\eqref{P1-multi} can be obtained via block-coordinate descent, where each \emph{block-variable} is cyclically optimized while leaving the others to their current value, such that the objective is nonincreasing with each iteration. Here, the natural block-variables are $\bm U$ and $\{\bm V_n\}_{n=1}^N$. Consider first the minimization over $\bm U$, and partition ${\bm V}_n$ as $[{\bm G}_{n} \ {\bm g}_{n}]\transp$, with $\bm G_n \in \mathbb C^{K_s\times Q}$ and $\bm g_n \in\mathbb C^{K_s}$, and $\bm Y_n$ as $[\bm Y_{n,1}; \bm Y_{n,2}]$, with $\bm Y_{n,1}\in \mathbb C^{P_n \times K_s}$ and $\bm Y_{n,2}\in \mathbb C^{(D_0+D_n) \times K_s}$. Then, exploiting \eqref{P1-multi-C2}, we have that
\begin{multline}\label{P1-multi-costraint}
	\Vert \bm Y_n - \bm{T}_n\bm{V}_n\Vert^2_F = \Vert \bm Y_{n,1} - \bm P_n{\bm G}_n\transp  - {\bm 1}_{P_n} {\bm g}_{n}\transp \Vert^2_F \\ + \Vert \bm Y_{n,2} - {\bm F}_n \bm U{\bm G}_n\transp - {\bm 1}_{D_0+D_n} {\bm g}_{n}\transp \Vert^2_F .
\end{multline}
Notice also that
\begin{multline}
\sum_{n=1}^N \Vert \bm Y_{n,2} - {\bm F}_n \bm U{\bm G}_n\transp - {\bm 1}_{D_0+D_n} {\bm g}_{n}\transp \Vert^2_F= \\  \sum_{n=1}^N \|\tilde{\bm y}_{n,2} - \tilde {\bm G}_n\bm u - \tilde{\bm g}_n\|^2 =  \|\tilde{\bm y}_2 - \tilde {\bm G}\bm u - \tilde{\bm g}\|^2, \label{obj_sub_U}
\end{multline}
where $\tilde{\bm y}_{n,2} = \mathrm{vec}\{\bm Y_{n,2}\}$, $\tilde{\bm y}_2 = [\tilde {\bm y}_{1,2}; \ \cdots; \ \tilde{\bm y}_{N,2}]$, $\bm u=\mathrm{vec}\{\bm U\}$, $\tilde {\bm G}_n = {\bm G}_n\otimes {\bm F}_n$, $\tilde {\bm G} = [\tilde {\bm G}_1; \ \cdots; \ \tilde {\bm G}_{N}]$, $ \tilde{\bm g}_n={\bm g}_{n}\otimes {\bm 1}_{D_0+D_n}$,  and $\tilde {\bm g} = [\tilde {\bm g}_1; \ \cdots; \ \tilde {\bm g}_{N}]$. At this point, from~\eqref{P1-multi-costraint} and~\eqref{obj_sub_U}, the problem to be solved here is
\begin{equation} 
 \min_{\bm u\in \mathcal X^{DQ}} \|\tilde{\bm y}_2 - \tilde{\bm G} \bm u- \tilde{\bm g}\|^2 + \lambda_{u}\|\bm u\|^2. \label{prob_u}
\end{equation}
Since $\mathcal X$ is a discrete alphabet, a closed-form solution to~\eqref{prob_u} is not available, and an exhaustive search is required. Consider now the minimization over $\{\bm V_n\}_{n=1}^N$, which results in $N$ decoupled  subproblems, namely,
\begin{equation}
\min_{\bm V_n \in \mathbb{C}^{(Q+1)\times K_s}} \|\bm Y_n - \bm T_n \bm V_n\|_F^2 + \lambda_{v,n}\|\bm V_n\|_F^2, 
\end{equation}
for $n=1,\ldots, N$; in particular, the $n$-th subproblem admits the following solution
\begin{equation}
\bm V_n^\star = (\bm T_n\herm \bm T_n + \lambda_{v,n} \bm I_{Q+1})^{-1} \bm T_n\herm \bm Y_n. \label{V_n_sol}
\end{equation}
The decoding procedure is summarized in Alg.~\ref{alg_1}:  we refer to it as the Alternating data Symbol and subChannel Estimation (ASCE). The computational cost per iteration is $\mathcal O\bigl(N(LQ^2 + Q^3 + K_sLQ +K_sQ^2)\bigr)$ in line~\ref{Alg_1_update_V} and $\mathcal O\bigl(K_s(NL-P)DQ |\mathcal X|^{DQ}\bigr)$ in line~\ref{Alg_1_update_U}. Since the objective function is bounded from below and not increased at each iteration, Alg.~\ref{alg_1} monotonically converges to a stationary point; however, the global optimality of this solution cannot be guaranteed. 
\begin{algorithm}[t]
 \caption{Proposed solution to Problem~\eqref{P1-multi} \label{alg_1}}
 \begin{algorithmic}[1]
 \STATE choose $\bm U\in\mathcal X^{D, Q+1}$
 \REPEAT
    \STATE update $\{\bm V_n\}_{n=1}^N$ with~\eqref{V_n_sol} \label{Alg_1_update_V}
    \STATE update $\bm U$ by solving~\eqref{prob_u} \label{Alg_1_update_U}
 \UNTIL convergence
 \RETURN $\bm U$, $\{\bm V_n\}_{n=1}^N$
 \end{algorithmic}
\end{algorithm}

When $DQ$ is large, the discrete search in~\eqref{prob_u} is prohibitive. In this case, we relax Problem~\eqref{prob_u} by enlarging the search set to $\mathbb C^{DQ}$; the corresponding solution is available in closed-form and is equal to
\begin{equation}
\mathrm{vec}\{\bm U^\star \} =  (\tilde{\bm G}\herm \tilde{\bm G} + \lambda_u \bm I_{DQ})^{-1}\tilde{\bm G}\herm (\tilde{\bm y} - \tilde{\bm g}), \label{U_rel_sol}
\end{equation}
thus entailing a computational cost $\mathcal O\bigl(D^3Q^3+K_s(NL-P)D^3Q^3\bigr)$. At convergence of the resulting block-coordinate descent procedure, we then perform symbol slicing on the returned $\bm U$ to estimate $\bm D$. We refer to this decoding procedure as the Relaxed ASCE (R-ASCE).

\subsection{Algorithm 2} \label{alg_2_sec}
We propose here to solve the following constrained regularized least squares problem 
\begin{subequations}\label{P4}
\begin{align}
& \hspace{-1.8cm} \min_{\substack{{\bm U} \in \mathcal{X}^{D\times Q}\\ \hspace{10pt}\bar{\bm{v}}_{q,n}\in \mathbb{C}^{\bar{K}}\,\forall q,n\\
 \hspace{45pt}\bar{d}_{q}\in\{0,\ldots,\lceil\tau_{h} K_s/T_s\rceil\}\, \forall q
\\ \bm{g}_{n}\in \mathbb{C}^{K_s}\, \forall n 
}} \sum_{n=1}^{N}\|\bm Y_n - \bm T_n {\bm V}_{n}\|_F^2  + \lambda_{u}\|\bm U\|_F^2  \notag \\[-30pt] & \hspace{3.5cm}+ \sum_{n=1}^{N}\lambda_{v,n}\|{\bm V}_n\|_F^2, \\[10pt]
& \text{s.t.}\;   \bm T_n=
    \begin{bmatrix}
        {\bm P}_{n} & {\bm 1}_{P_n} \\ 
        {\bm F}_{n} {\bm U} & {\bm 1}_{D_0+D_n}
    \end{bmatrix}, \, \forall n, \label{P4_C_T}
\\
& \hphantom{\text{s.t.}\;} \bm{V}_n=\left[\bm{S}(\bar{d}_{1})\bar{\bm{v}}_{1,n} \ \cdots \ \bm{S}(\bar{d}_{Q}) \bar{\bm{v}}_{Q,n} \ \bm{g}_n\right]\transp, \, \forall n.\label{P4_C_V}
\end{align}%
\end{subequations}
Differently from Problem~\eqref{P1-multi}, the constraint in~\eqref{P4_C_T} now exploits the block structure in the first $Q$ rows of the subchannel response matrices that is implied by~\eqref{A-structure}. To be more specific, while Problem~\eqref{P1-multi} attempts to recover ${\bm V}_{n}$ as a whole, Problem~\eqref{P4}  attempts to recover only the non-zero elements of ${\bm V}_{n}$; intuitively, this latter approach  can potentially give better performance at the price of introducing a discrete search over the unknown delay offsets of the tags.

We resort again to the block-coordinate descent method and cyclically optimize over the block-variables $\bm U$, $\bigl\{d_1, \{ \bar{\bm v}_{1,n}\}_{n=1}^N\bigr\},\ldots, \bigl\{d_Q, \{ \bar{\bm v}_{Q,n}\}_{n=1}^N \bigr\}$, and $\{\bm g_n\}_{n=1}^N$. The minimization over $\bm U$ can again be carried out by solving Problem~\eqref{prob_u}. As to the minimization over $\bigl\{d_q, \{\bm v_{q,n}\}_{n=1}^N\bigr\}$, for $q=1,\ldots,Q$, the problem to be solved is
\begin{multline}
 \min_{\substack{\bar{\bm v}_{q,n}\in \mathbb{C}^{\bar{K}}, \, \forall n\\ \bar{d}_q\in\{0,\ldots,\lceil\tau_{h} K_s/T_s\rceil\}}} \sum_{n=1}^{N}\Bigg\| \bm Y_n - \sum_{\substack{j=1\\ j\neq q}}^Q [\bm{T}_n]_{:,j} \left(\bm{S}(\bar{d}_j)\bar{\bm{v}}_{j,n}\right)\transp \\  -  [\bm{T}_n]_{:,q} \left(\bm{S}(\bar{d}_q)\bar{\bm{v}}_{q,n}\right)\transp  - \bm{1}_{L} \bm{g}_n \transp\Bigg\|_F^2 + \sum_{n=1}^{N} \lambda_{v,n}\|\bm{S}(\bar{d}_q)\bar{\bm{v}}_{q,n}\|^2. \label{max_vd_alg_2}
\end{multline}
To proceed, notice that $\|\bm{S}(\bar{d}_q)\bar{\bm{v}}_{q,n}\|^2 = \|\bar{\bm{v}}_{q,n}\|^2$; also, upon defining
\begin{equation}
\bar {\bm{Y}}_n = \bm Y_n - \sum_{\substack{j=1\\ j\neq q}}^Q [\bm{T}_n]_{:,j} \left(\bm{S}(\bar{d}_j)\bar{\bm{v}}_{j,n}\right)\transp   - \bm{1}_{L} \bm{g}_n \transp,
\end{equation}
we have that the Frobenius norm in the objective function of~\eqref{max_vd_alg_2} can written as
\begin{multline}
  \left\|\bar{\bm Y}_n - [\bm{T}_n]_{:,q}\left(\bm{S}(\bar{d}_q)\bar{\bm{v}}_{q,n}\right)\transp\right\|_F^2
   =\left\|[\bar{\bm Y_n}]_{:,1:\bar d_{q,n}} \right\|_F^2  \\
 + \left\|[\bar{\bm Y}_n]_{:, \bar d_q+1: \bar d_q + \bar K} - [\bm{T}_n]_{:,q} \bar{\bm{v}}_{q,n}\transp \right\|_F^2 + \left\|[\bar{\bm Y_n}]_{:,\bar d_{q,n}+\bar K +1:K_s} \right\|_F^2.
\end{multline}
Therefore, Problem~\eqref{max_vd_alg_2} reduces to
\begin{align}
  \min_{\substack{\bar{\bm v}_{q,n}\in \mathbb{C}^{\bar{K}}, \, \forall n\\ \bar{d}_q\in\{0,\ldots,\lceil\tau_{h} K_s/T_s\rceil\}}} f \left(\left\{\bar{\bm v}_{q,n}\right\}_{n=1}^N, \bar{d}_q \right),
  \label{delay_est_eq}
\end{align}
where
\begin{align}
  f \left(\left\{\bar{\bm v}_{q,n}\right\}_{n=1}^N, \bar{d}_q \right)&= \sum_{n=1}^{N}\left\|[\bar{\bm Y}_n]_{:, \bar d_q+1: \bar d_q + \bar K}  - [\bm{T}_n]_{:,q} \bar{\bm{v}}_{q,n}\transp \right\|_F^2 \notag \\ &\quad +  \sum_{n=1}^{N} \lambda_{v,n}\|\bar{\bm{v}}_{q,n}\|^2.
\end{align}
For fixed $\bar{d}_q$, the objective function in~\eqref{delay_est_eq} is minimized by
\begin{equation}
 \bar{\bm v}_{q,n}\transp (\bar{d}_q) = \frac{ [\bm{T}_n]_{:,q}\herm [\bar{\bm Y}_n]_{:, \bar d_q+1: \bar d_q + \bar K}} {\lambda_{v,n} + \|[\bm{T}_n]_{:,q} \|^2}.
\label{vq_given_dq}
\end{equation}
Hence, the solution to Problem~\eqref{delay_est_eq} is
\begin{subequations}
 \begin{align}
 \bar{d}_{q}^\star  &=  \argmin_{\bar{d}_q\in\{0,\ldots,\lceil\tau_{h} K_s/T_s\rceil\}} f \left(\left\{\bar{\bm v}_{q,n}(\bar d_q)\right\}_{n=1}^N, \bar{d}_q\right), \label{dqbar} \\
 \bar{\bm v}_{q,n}^\star &=  \bar{\bm v}_q(\bar{d}_{q}^\star).\label{vqbar}
 \end{align} \label{dqbar_vqbar} 
\end{subequations}
Finally, the minimization over $\{\bm g_n\}_{n=1}^N$ decouples in $N$ subproblems, namely,
\begin{equation}
 \min_{\bm g_n \in \mathbb{C}^{K_s}} \bigg\|\bm Y_n - \sum_{j=1}^Q [\bm{T}_n]_{:,j} \left(\bm{S}(\bar{d}_j)\bar{\bm{v}}_{j,n}\right)\transp -\bm{1}_{L} \bm g_n\transp\bigg\|_F^2 + \lambda_{v,n}\| \bm g_n\|^2, \label{min_g}
\end{equation}
for $n=1,\ldots, N$, and the solution to the $n$-th subproblem is
\begin{equation}
  \bm g_n\transp =\frac{\bm{1}_{L}\transp \left(\bm Y_n - \sum_{j=1}^Q [\bm{T}_n]_{:,j} \left(\bm{S}(\bar{d}_j)\bar{\bm{v}}_{j,n}\right)\transp \right)}{L + \lambda_{v,n} }. \label{sol_g_n}
\end{equation}
The decoding procedure is summarized in Alg.~\ref{alg_2}: we refer to it as the ASCE with Delay recovery (ASCE-D). The computational cost per iteration is $\mathcal O(LNQ^2 K_s^2)$ in lines~\ref{Alg_2_update_d_v_start}--\ref{Alg_2_update_d_v_end}, $\mathcal O(LNQK_s)$ in line~\ref{Alg_2_update_g}, and $\mathcal O\bigl(K_s(NL-P)DQ |\mathcal X|^{DQ}\bigr)$ in line~\ref{Alg_2_update_U}. Since the objective function is bounded from below and not increased at each iteration, Alg.~\ref{alg_2} monotonically converges to a stationary point; however, the global optimality of this solution cannot be guaranteed.
 	
As in Sec.~\ref{alg_1_sec}, the update of $\bm U$ in line~\ref{Alg_2_update_U} of Alg.~\ref{alg_2} is done using~\eqref{U_rel_sol} when $DQ$ is large; at convergence of the block-coordinate descent procedure, symbol slicing on the returned $\bm U$ is performed. We refer to this decoding procedure as the Relaxed ASCE-D (R-ASCE-D). 
\begin{algorithm}[t]
 \caption{Proposed solution to Problem~\eqref{P4} \label{alg_2}}
 \begin{algorithmic}[1]
 \STATE choose $\bm U\in\mathcal X^{D\times Q+1}$
 \REPEAT
    \FOR{$q=1,\ldots,Q$}\label{Alg_2_update_d_v_start}
        \STATE update $\bigl\{ \bar d_q, \{\bar{\bm v}_{q,n}\}_{n=1}^N \bigr\}$ with~\eqref{dqbar_vqbar} \label{Alg_2_update_d_v}
    \ENDFOR\label{Alg_2_update_d_v_end} 
    \STATE update $\{{\bm g}_n \}_{n=1}^N$ with~\eqref{sol_g_n} \label{Alg_2_update_g}
    \STATE update $\bm U$ by solving~\eqref{prob_u} \label{Alg_2_update_U}
 \UNTIL convergence
 \STATE $\bm{V}_n=\left[\bm{S}(\bar{d}_{1})\bar{\bm{v}}_{1,n} \ \cdots \ \bm{S}(\bar{d}_{Q}) \bar{\bm{v}}_{Q,n} \ \bm{g}_n\right]\transp$, $\forall n$
 \RETURN $\bm U$, $\{\bm V_n\}_{n=1}^N$
 \end{algorithmic}
\end{algorithm}

\subsection{Missing measurements}
\label{sec:missing_data}
The previous algorithms can be modified to handle missing entries in $\{\bm Y_n\}_{n=1}^{N}$, which results in a matrix completion problem~\cite{Nguyen-2019}. For example, this is the case when a sub-Nyquist sampling rate is employed to reduce the complexity of the reader; in addition, the reader may discard the measurements corrupted by strong radar interference: indeed, in this case, the received signal may saturate the low-noise amplifier, which in turn may cause a severe overload distortion at the output of the analog-to-digital converter. For brevity, we only show next how the decoding algorithm in Sec.~\ref{alg_2_sec} can be modified. 

To proceed, let $\bm E_n \in \{0,1\}^{L\times K_s}$ be defined as
\begin{equation}
[\bm E_n]_{i,j} = \begin{cases}
    1, & \text{if $[\bm Y_n]_{i,j}$ is observed},\\
    0, & \text{otherwise.}
\end{cases}
\end{equation}
Then, the problem in~\eqref{P4} can be modified as
\begin{align}\label{P_missing_data}
& \hspace{-2.1cm} \min_{\substack{{\bm U} \in \mathcal{X}^{D\times Q}\\ \hspace{10pt}\bar{\bm{v}}_{q,n}\in \mathbb{C}^{\bar{K}}\,\forall q,n\\
 \hspace{45pt}\bar{d}_{q}\in\{0,\ldots,\lceil\tau_{h} K_s/T_s\rceil\}\, \forall q
\\ \bm{g}_{n}\in \mathbb{C}^{K_s}\, \forall n 
}} \sum_{n=1}^{N} \left\|\bm E_n \! \odot \! (\bm Y_n \!-\!\bm T_n {\bm V}_{n}) \right\|_F^2 \notag \\[-25 pt] 
&\hspace{3cm}+ \lambda_{u}\|\bm U\|_F^2 + \sum_{n=1}^{N}\lambda_{v,n}\|{\bm V}_n\|_F^2, \\[10 pt]
& \text{s.t.}\; \eqref{P4_C_T} \text{ and } \eqref{P4_C_V}. \notag
\end{align}
In other words, we ignore in the objective function the missing entries and fit all search variables according only to the observed entries of $\bm Y_1,\ldots,\bm Y_N$.

For the minimization over $\bm U$, we proceed as in Sec.~\ref{alg_1_sec}. Specifically, let $\bm E_{n,2}\in \mathbb C^{(D_0+D_n) \times K_s}$ be the matrix obtained by taking the last $D_0+D_n$ rows of $\bm E_n$, and let $\tilde{\bm e}_{n,2} = \mathrm{vec}\{\bm E_{n,2}\}$ and $\tilde{\bm e}_2 = [\tilde {\bm e}_{2,1}; \ \cdots; \ \tilde{\bm e}_{2,N}]$. Then, the problem to be solved is
\begin{equation} 
 \min_{\bm u\in \mathcal X^{DQ}} \bigl\| \tilde{\bm e}_2 \odot (\tilde{\bm y}_2 -  \tilde{\bm G} \bm u- \tilde{\bm g} ) \bigr\|^2 + \lambda_{u}\|\bm u\|^2. \label{als-missing-t-update}
\end{equation}
Again, this problem can be solved by exhaustive search, or it can be relaxed; in this latter case, the closed-form solution is 
\begin{align}
 \mathrm{vec}\{\bm U^\star \} &=  (\tilde{\bm G}\herm \mathrm{diag} (\tilde{\bm e}_2)\tilde{\bm G} + \lambda_u \bm I_{DQ})^{-1}\notag \\ &\quad \times \tilde{\bm G}\herm \mathrm{diag} (\tilde{\bm e}_2) (\tilde{\bm y} - \tilde{\bm g}). 
\end{align}

As to the minimization over $\bigl\{d_q, \{\bm v_{q,n}\}_{n=1}^N\bigr\}$, for $q=1,\ldots,Q$, proceeding as in Sec.~\ref{alg_2_sec}, we arrive at the following modified version of Problem~\eqref{delay_est_eq}
\begin{equation}
  \min_{\substack{\bar{\bm v}_{q,n}\in \mathbb{C}^{\bar{K}}, \, \forall n\\ \bar{d}_q\in\{0,\ldots,\lceil\tau_{h} K_s/T_s\rceil\}}}  f_\text{miss} \left(\left\{\bar{\bm v}_{q,n}\right\}_{n=1}^N, \bar{d}_q \right), \label{delay_est_eq_missing}
\end{equation}
where
\begin{multline}
    f_\text{miss} \left(\left\{\bar{\bm v}_{q,n}\right\}_{n=1}^N, \bar{d}_q \right) = \sum_{n=1}^{N}\Bigr\|  [\bm E_n]_{:, \bar d_q+1: \bar d_q + \bar K} \\ \!\odot\! \left(\![\bar{\bm Y}_n]_{:, \bar d_q+1: \bar d_q + \bar K} - [\bm{T}_n]_{:,q} \bar{\bm{v}}_{q,n}\transp\!\right) \!\!\Big\|_F^2
    \!+\!  \sum_{n=1}^{N} \!\lambda_{v,n}\|\bar{\bm{v}}_{q,n}\|^2\!.
\end{multline}
For fixed $\bar{d}_q$, the objective function in~\eqref{delay_est_eq_missing} is minimized by
\begin{multline}
 \bar{\bm v}_{q,n}\transp (\bar{d}_q) = [\bm{T}_n]_{:,q}\herm \left( \bm [\bm E_n \odot \bar{\bm Y}_n]_{:, \bar d_q+1: \bar d_q + \bar K} \right)  \\ 
 \times  \mathrm{diag} \Biggl( \Biggl\{ \frac{1}{\lambda_{v,n} + \| [\bm E_n]_{:, \bar d_q+k} \odot [\bm{T}_n]_{:,q} \|^2} \Biggr\}_{k=1}^{\bar K}\Biggr).
\label{vq_given_dq_missing}
\end{multline}
Therefore, the solution to Problem~\eqref{delay_est_eq_missing} is
\begin{subequations}
 \begin{align}
 \bar{d}_{q}^\star  &=  \argmin_{\substack{\bar{\bm v}_{q,n}\in \mathbb{C}^{\bar{K}}, \, \forall n\\ \bar{d}_q\in\{0,\ldots,\lceil\tau_{h} K_s/T_s\rceil\}}} \!\!\! f_\text{miss} \left(\left\{\bar{\bm v}_{q,n}(\bar{d}_q)\right\}_{n=1}^N, \bar{d}_q \right), \\
 \bar{\bm v}_{q,n}^\star &=  \bar{\bm v}_q(\bar{d}_{q}^\star).
 \end{align}
\end{subequations}
Finally, the minimization over $\{\bm g_n\}_{n=1}^N$ in~\eqref{min_g} becomes
\begin{multline}
 \min_{\bm g_n \in \mathbb{C}^{K_s}} \Biggl\| \bm E_n \odot \Biggl( \bm Y_n -  \sum_{j=1}^Q [\bm{T}_n]_{:,j} \left(\bm{S}(\bar{d}_j)\bar{\bm{v}}_{j,n}\right)\transp -\bm{1}_{L} \bm g_n\transp \Biggr) \Biggr\|_F^2 \\+ \lambda_{v,n}\| \bm g_n\|^2,
\end{multline}
for $n=1,\ldots, N$, and it is not difficult to show that the solution to the $n$-th subproblem is
\begin{align}
  \bm g_n\transp &=  \bm 1_L\transp \Biggl(\bm E_n \odot \Biggl( \bm Y_n - \sum_{j=1}^Q [\bm{T}_n]_{:,j} \left(\bm{S}(\bar{d}_j)\bar{\bm{v}}_{j,n}\right)\transp \Biggr)\Biggr) \notag\\ &\quad \times \mathrm{diag} \Biggl(\left\{ \frac{1}{[\bm 1_L\transp \bm E_n]_k + \lambda_{v,n} } \right\}_{k=1}^{K_s}\Biggr) .
\end{align}

\section{Numerical Analysis}\label{SEC:Numerical analysis}
Here, we assess the performance of the proposed encoding/decoding strategies. The entries of the noise matrices are modeled as independent circularly-symmetric Gaussian random variables with variance $\sigma^{2}_{w}$. As to the tags, an $M$-PSK alphabet with a Gray bit-to-symbol mapping is considered, while the delay offset is randomly chosen from $\{0,\ldots, K_{s}-\bar{K}\}$. Following~\cite{Shnidman-1999,2022-Venturino-ABC}, the entries of $\bar{\bm{\alpha}}_{q,n}$ are generated as the sum of a specular and a diffuse component, namely, $\bar{\bm{\alpha}}_{q,n} = \sigma_{\alpha,\text{spe}}\e^{\i 2 \pi \xi_{q,n}}\bm{1}_{\bar K} + \bar{\bm{\alpha}}_{\text{dif},q,n}$, where $\sigma_{\alpha,\text{spe}}$ is a positive deterministic quantity, $ \xi_{q,n}$ is a random phase, and $\bar{\bm{\alpha}}_{\text{dif},q,n}$ is a circularly-symmetric Gaussian vector with covariance matrix $\sigma_{\alpha,\text{dif}}^2 \bm I_{\bar K}$; accordingly, $\kappa_{\alpha}=\sigma^2_{\alpha,\text{spe}}/\sigma^2_{\alpha,\text{dif}} $ is the power ratio between the specular and diffuse components, while $\text{SNR}=(\kappa_{\alpha}+1) \sigma_{\alpha,\text{dif}}^2 / \sigma_w^2$ is the signal-to-noise ratio. A similar model is employed for $\bm{i}_{n}$; in this latter case, the power ratio between the specular and diffuse components is denoted by $\kappa_{i}=\sigma^2_{i,\text{spe}}/\sigma^2_{i,\text{dif}}$, while $\text{INR}=(\kappa_{i}+1) \sigma_{i,\text{dif}}^2 / \sigma_w^2$  is the interference-to-noise ratio and $\text{SIR}=\text{SNR}/\text{INR}$ is the signal-to-interference ratio. 

The system performance is assessed in terms of the BER averaged over all tags and the NRMSE in the estimation of the subchannel response matrices defined as
\begin{equation}
\mathrm{NRMSE}= \sqrt{\frac{\mathrm{E}\left[\sum_{n=1}^{N}\|\bm A_n -\hat{\bm A}_n\|_F^2\right]}{\mathrm{E}\left[\sum_{n=1}^{N}\|\bm A_n\|_F^2\right]}}.
\end{equation}
where $\{\hat{\bm A}_n\}_{n=1}^{N}$ are the estimated subchannel response matrices,  and $\mathrm{E}\bigl[\sum_{n=1}^{N}\|\bm A_n\|_F^2 \bigr] = N Q \bar K (\sigma^2_{\alpha,\text{spe}}+\sigma^2_{\alpha,\text{diff}}) + NK_s (\sigma^2_{i,\text{spe}}+\sigma^2_{i,\text{diff}}) $. We set the regularization parameters as $\lambda_u = \sigma^2_w$ and $\lambda_{v,n} = 0.1$, for $n=1,\ldots,N$; also, we set $\kappa_{\alpha}=\kappa_{i}=-10$~dB, $\text{SIR}=-10$~dB, $K_{s}=8$, $\bar K=3$, and, unless otherwise stated, $M=2$ and $Q=2$. All iterative algorithms are stopped when  $|f_{t+1} - f_t|  < 10^{-8} {f_t}$, where $f_t$ is the value of the objective function at iteration $t$.

\subsection{Encoding on a Single Subchannel}
Consider here the use of a single subchannel, i.e., $N=1$: in this case, Theorem~\ref{thm:unique} implies that noiseless unique recovery is possible with at least $Q+1$ pilot symbols.

\begin{figure}
	\centering
	\includegraphics[width=0.48\textwidth,trim=0cm 0.2cm 0cm 0.3cm, clip]{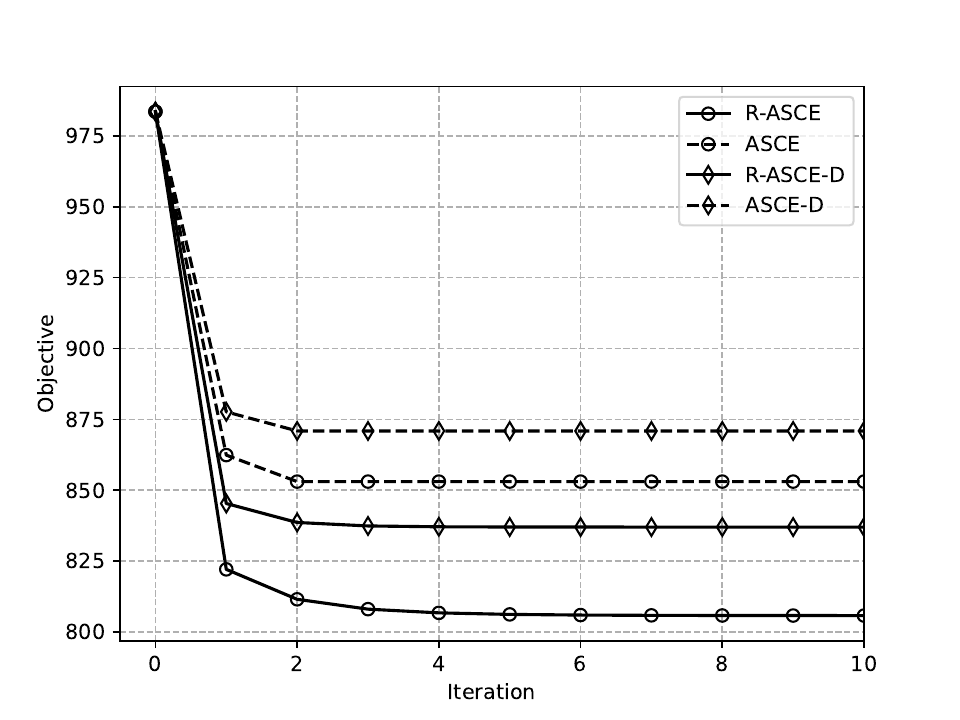}
	\caption{Objective function in~\eqref{P1-multi} and~\eqref{P4} versus iterations when $N=1$, $L=8$, $P_1=4$, and $\text{SNR}=10$~dB. Other system parameters: $R=0.5$ bits/subchannel-use/tag, $\text{SIR}=-10$~dB, $M=2$, $Q=2$, $K_s=8$, and $\bar K=3$.}
	\label{convergence_plot}
\end{figure}

\begin{figure}[!t]
	\centering
	\includegraphics[width=0.44\textwidth,trim=0cm 0.4cm 0cm 0.3cm, clip]{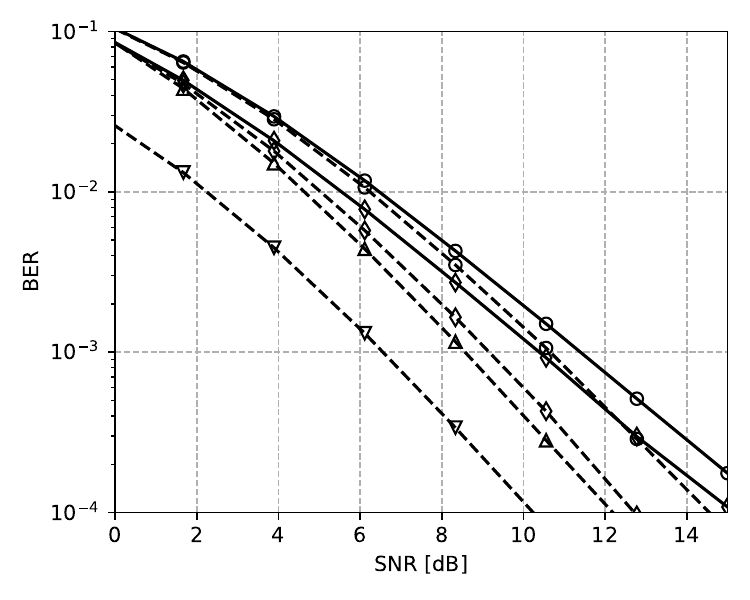}
 \\[0.3cm]
	\includegraphics[width=0.44\textwidth,trim=0cm 0.4cm 0cm 0.3cm, clip]{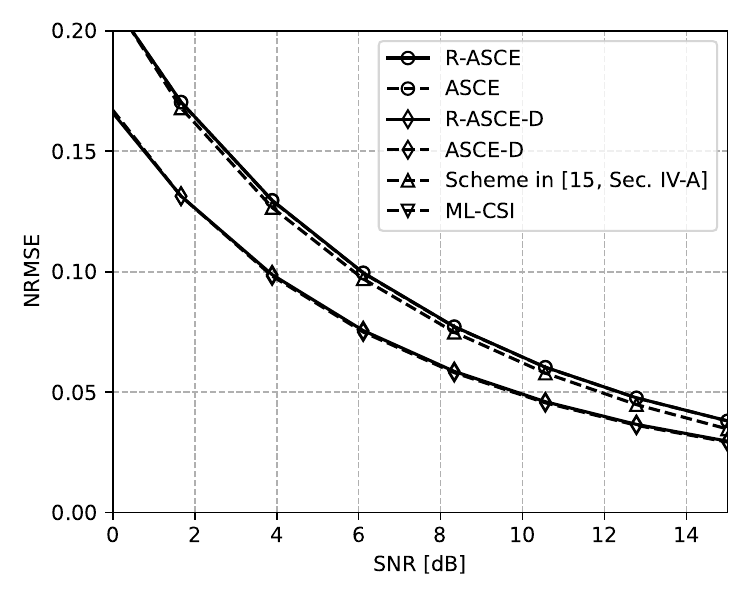}
	\caption{BER (top) and NRMSE (bottom)  versus SNR when $N=1$, $L=8$, and $P_1=4$. Other system parameters: $R=0.5$ bits/subchannel-use/tag, $\text{SIR}=-10$~dB, $M=2$, $Q=2$, $K_s=8$, and $\bar K=3$.}
	\label{BER_NRMSE_vs_SNR}
\end{figure}

Assume first $L=8$ and $P_1=4$, which gives a transmission rate of $R=0.5$ bits/subchannel-use/tag. For the proposed decoding algorithms, Fig.~\ref{convergence_plot} shows the value of the objective function in~\eqref{P1-multi} and~\eqref{P4} versus the iteration number for a single sample optimization trajectory when $\text{SNR}=10$~dB. Also, Fig.~\ref{BER_NRMSE_vs_SNR} reports BER and NRMSE versus SNR. As benchmarks, Fig.~\ref{BER_NRMSE_vs_SNR} includes the BER for the proposed encoding strategy when the ML decoder with perfect CSI (ML-CSI) is employed; also, for the same transmission rate, it includes the BER and NRMSE of the encoding/decoding scheme in~\cite[Sec. IV-A]{2022-Venturino-ABC}. The following remarks are now in order. 

\begin{figure}[!t]
	\centering
	\includegraphics[width=0.44\textwidth,trim=0cm 0.4cm 0cm 0.3cm, clip]{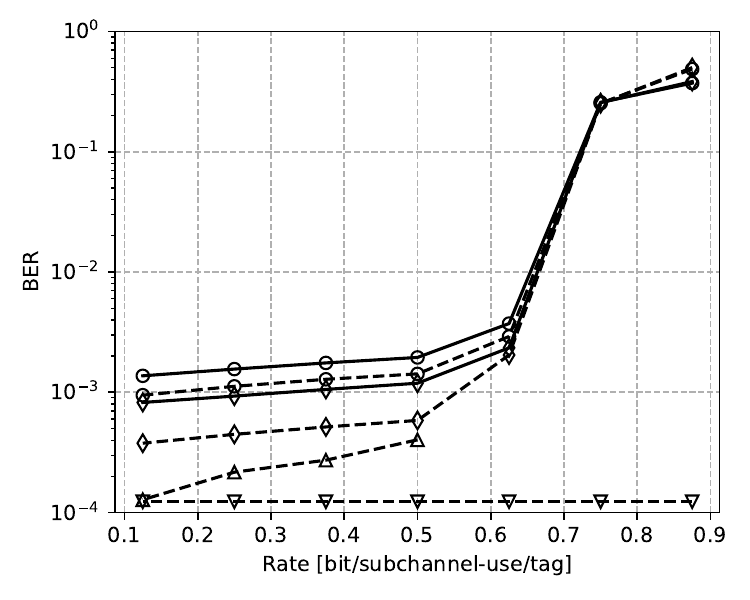}
 \\[0.3cm]
	\includegraphics[width=0.44\textwidth,trim=0cm 0.4cm 0cm 0.3cm, clip]{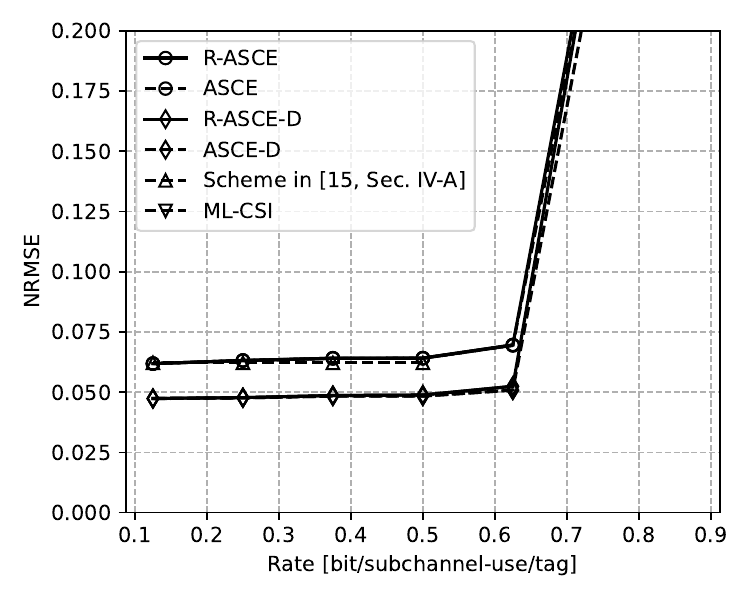}
	\caption{BER (top) and NRMSE (bottom) versus the transmission rate when $N=1$, $L=8$, and  $P_1$ is varied from $7$ (lowest rate) to $1$ (highest rate). Other system parameters: $\text{SNR}=10$~dB, $\text{SIR}=-10$~dB, $M=2$, $Q=2$, $K_s=8$, $\bar K=3$.}
	\label{BER_NRMSE_vs_P_fixed_L}
\end{figure}

\begin{itemize}
\item As anticipated in Sec.~\ref{SEC:algorithms}, it is seen from Fig.~\ref{convergence_plot} that the proposed decoding algorithms are monotonically convergent; also, they present a similar convergence speed, and less than 10 iterations are usually sufficient.

\item At convergence, R-ASCE-D provides a higher value of the objective function than R-ASCE. Indeed, constraining the channel structure reduces model over-fitting in R-ASCE-D: remarkably, this reflects in both a lower BER and a lower NRMSE, as seen from Fig.~\ref{BER_NRMSE_vs_SNR}. A similar conclusion holds when comparing  ASCE-D and ASCE.

\item At convergence, ASCE-D provides a higher value of the objective function than R-ASCE-D. Indeed, constraining the set of codewords reduces model over-fitting in ASCE-D: this reflects in a lower BER, while it has a negligible impact on the RMSE, as seen from Fig.~\ref{BER_NRMSE_vs_SNR}. A similar conclusion holds when comparing ASCE and R-ASCE.

\item ASCE-D and ASCE solve Problem~\eqref{prob_u} via an exhaustive search: this is possible here since $DQ=16$, but it will be infeasible in many of the following examples. Remarkably, the corresponding relaxed implementations incur a limited loss while entailing a computational cost that scales cubically (rather than exponentially) with $DQ$.

\item In Fig.~\ref{BER_NRMSE_vs_SNR} (top), the BER curves of ASCE and R-ASCE-D cross around $\text{SNR}=12$~dB: the understanding is that, at lower SNR values, constraining the set of codewords in ASCE is less rewarding than exploiting the additional structure of the subchannel response matrices in R-ASCE-D; at higher SNRs, the opposite occurs.

\item In Fig.~\ref{BER_NRMSE_vs_SNR} (top), ASCE-D presents an SNR gap of $2.5$~dB with respect to the ML decoder with CSI at $\text{BER}=10^{-3}$, which is a relevant result considering that only $4$ pilot symbols are employed here to enable signal recovery.

\item In Fig.~\ref{BER_NRMSE_vs_SNR} (top), the encoding/decoding scheme in~\cite{2022-Venturino-ABC} presents a better performance compared to the proposed solutions, with a SNR gap that depends on the employed decoding algorithm and ranging from about $0.5$~dB for ASCE-D to about $4$~dB for R-ASCE at $\text{BER}=10^{-3}$. This slight superiority may be explained because the scheme in~\cite{2022-Venturino-ABC} employs codewords orthogonal to the radar interference and ML decoding, which is not our case. Nevertheless, the following facts make the proposed encoding rule coupled with R-ASCE-D or R-ASCE decoding preferable in practice.
\begin{itemize}
	\item Differently from the proposed one, the previous encoder is hardly scalable with $Q$ and $D$; indeed, the codebooks of the tags are constructed via an exhaustive search to match the Conditions (P1s) and (P2s) given in~\cite[Sec.~IV-A]{2022-Venturino-ABC} and then stored into a look-up memory at the tag and reader. 
	
	\item The former encoding strategy supports transmission rates lower than those achievable by the proposed method (more on this when discussing Fig.~\ref{BER_NRMSE_vs_P_fixed_L}).
	
	\item Differently from R-ASCE-D and R-ASCE, the ML decoder in~\cite{2022-Venturino-ABC} is hardly scalable with $Q$ and $D$; indeed, it involves a computational cost that scales exponentially with $DQ$, i.e., $\mathcal O\bigl((LQ^2 + Q^3 + K_sLQ) |\mathcal X|^{DQ}\bigr)$.
	
	\item The encoding/decoding scheme in~\cite{2022-Venturino-ABC} can be extended to $N>1$ only if a plain repetition coding is performed across the subchannels (see Example~5 in~\cite{2022-Venturino-ABC}).
\end{itemize}
\end{itemize}

Next, Fig.~\ref{BER_NRMSE_vs_P_fixed_L} reports BER and NRMSE  of the proposed solutions versus the transmission rate when the number of pilot symbols $P_1$ is varied from $7$ (lowest rate) to $1$ (highest rate), $\text{SNR}=10$~dB, and $L=8$. Communication fails for rates larger than $0.625$ bits/subchannel-use/tag (corresponding to $P_1=3$), while BER and NRMSE remain stable at lower rates, in keeping with Theorem~\ref{thm:unique}; on the other hand, the encoding scheme in~\cite{2022-Venturino-ABC} cannot provide rates larger than $R=0.5109$ bits/subchannel-use/tag. To shed more light on this point, Table~\ref{tab_largest_rate} compares the largest transmission rate satisfying the hypotheses of Theorem~\ref{thm:unique} for the proposed encoding scheme with the largest transmission rate satisfying Conditions (P1s) and (P2s) in~\cite[Sec.~IV-A]{2022-Venturino-ABC}, for different $Q$ and $L$. It is seen that the former is always larger than the latter. 

\begin{table}[!t]
	\centering 
	\caption{Largest transmission rate in bits/subchannel-use/tag when $N=1$ and $M=2$. \label{tab_largest_rate}}
	\begin{tabular}{ccccc}
		\toprule
		\multirow{2}[1]{*}{$L$} & \multicolumn{2}{c}{$Q=1$} & \multicolumn{2}{c}{$Q=2$} \\
		\cmidrule(lr){2-3} \cmidrule(lr){4-5} & proposed & \cite[Sec. IV-A]{2022-Venturino-ABC}  & proposed & \cite[Sec. IV-A]{2022-Venturino-ABC} \\
		\midrule
		4 & 0.5 & 0.3962 & 0.25  & $-$  \\
		6 & 0.6667 & 0.5537 & 0.5 &  0.3870 \\
		8 & 0.75 & 0.6412 &  0.625 & 0.5109  \\
		10 & 0.8 & 0.6977 & 0.7  &  05977 \\
		12 & 0.8333 & 0.7907 & 0.75 &  0.6543 \\
		14 & 0.8571 & 0.7675 & 0.7857 & 0.6961 \\
		16 & 0.875& 0.7729 & 0.8125 &  0.7282\\
		18 & 0.889 & 0.8094 & 0.8333 & 0.7538  \\
		\bottomrule
	\end{tabular}
\end{table}

\begin{figure}[!t]
	\centering
	\includegraphics[width=0.44\textwidth,trim=0cm 0.4cm 0cm 0.3cm, clip]{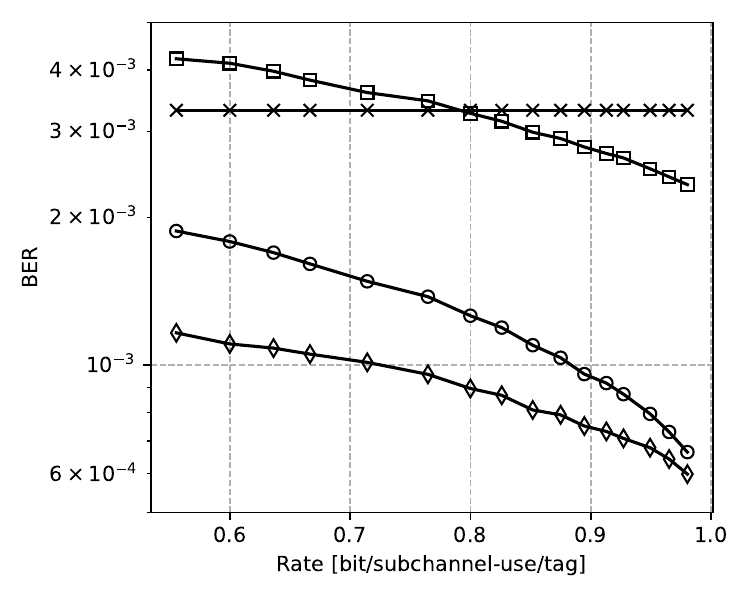}
     \\[0.3cm]
	\includegraphics[width=0.44\textwidth,trim=0cm 0.4cm 0cm 0.3cm, clip]{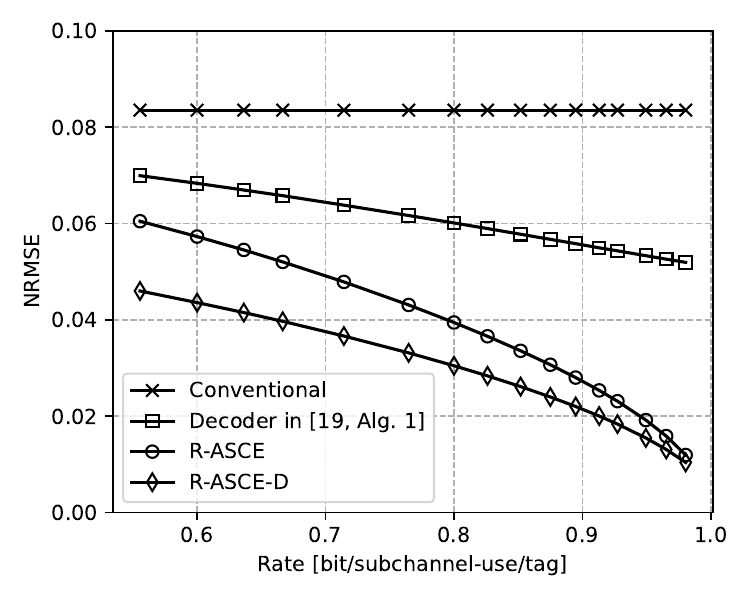}
    \caption{BER (top) and NRMSE (bottom)  versus the transmission rate when $N=1$, $P_1=4$, and  $D_1$ is varied from $5$ (lowest rate) to $200$ (highest rate). Other system parameters: $\text{SNR}=10$~dB, $\text{SIR}=-10$~dB,  $M=2$, $Q=2$, $K_s=8$, and $\bar K=3$.}
	\label{BER_NRMSE_vs_D_fixed_P} 
\end{figure}
Fig.~\ref{BER_NRMSE_vs_D_fixed_P} reports BER and NRMSE  of R-ASCE and R-ASCE-D versus the transmission rate when the number of data symbols $D_1$ is varied from $5$ (lowest rate) to $200$ (highest rate), $\text{SNR}=10$~dB, and $P_1=4$. As a benchmark, we include the performance of the semi-blind decoding strategy in \cite[Alg.~1]{8807374} and\footnote{To be more specific, the following problem
\begin{equation*}
 \underset{\substack{{\bm T}_1\in\mathbb{C}^{L\times (Q+1)}\\ \bm V_1 \in \mathbb{C}^{(Q+1)\times K_s}}}{\min} \|\bm Y_1 - \bm T_1 \bm V_1\|_F^2 + \lambda_{u}\|\bm T_1\|_F^2 +  \lambda_{v,1}\|\bm V_1\|_F^2
\end{equation*}
is suboptimally solved via an alternating least squares procedure. At each iteration, $\bm V_1$ is updated as in~\eqref{V_n_sol} with a computational cost $\mathcal{O}\bigl(LQ^2 + Q^3 + K_sLQ +K_sQ^2\bigr)$; instead,
$\bm T_1$ is updated as $\bm Y_1 \bm V_1\herm(\bm V_1\bm V_1\herm+ \lambda_{u} \bm I_{Q+1})^{-1}$ with a computational cost $\mathcal O\bigl(LQ^2 + Q^3 + K_sLQ +K_sQ^2\bigr)$. At convergence, the knowledge of the pilot symbols is exploited to remove the ambiguity in the recovered decomposition. Partition $\bm{T}_1$ as $[ \bm{T}_{1,p};  \bm{T}_{1,d}]$, where $\bm{T}_{1,p}\in\mathbb{C}^{P_{1}\times (Q+1)}$ and  $\bm{T}_{1,d}\in\mathbb{C}^{D_{1}\times (Q+1)}$; then, an estimate of the data symbol matrix and of the subchannel response matrix are $[\bm{T}_{1,d} {\bm R}_1^{-1}]_{:,1:Q}$  and ${\bm R}_1\bm{V}_{1} $, respectively, where $\bm{R}_1^{-1}= \bigl(\bm{T}_{1,p}\bigr)^{\dag} [ \bm{P}_1 \ \bm{1}_{P_1} ]$; finally, symbol slicing is performed on the entries of $[\bm{T}_{1,d} {\bm R}_1^{-1}]_{:,1:Q}$.}  of a conventional decoding method that first uses a linear minimum mean square error (LMMSE) filter to estimate the subchannel response matrix using the pilot symbols, then uses a LMMSE filter to obtain an estimate of the data symbols based the estimated subchannel response matrix, and final performs symbol slicing on the returned soft data symbols.\footnote{In conventional decoding, the LMMSE-based estimation of the subchannel response and the data symbols has a computational cost $\mathcal O\bigl(P_1Q^2 + Q^3 + K_sP_1Q +K_sQ^2\bigr)$ and $\mathcal O\bigl(D_1Q^2 + Q^3 + K_sD_1Q +K_sQ^2\bigr)$, respectively.}  While the performance of the conventional method remains the same as $D_1$ increases, the performance of all other solutions improves since they utilize the data symbols as \emph{virtual pilot symbols} to improve signal recovery. Our solutions significantly outperform the previous method in~\cite[Alg.~1]{8807374}, as they exploit the specific structure of the matrix factors induced by
the encoding rule and by the timing of the tags in each iteration of the alternating optimization.

\begin{figure}
\centering 
\includegraphics[width=0.44\textwidth,trim=0cm 0.4cm 0cm 0.3cm, clip]{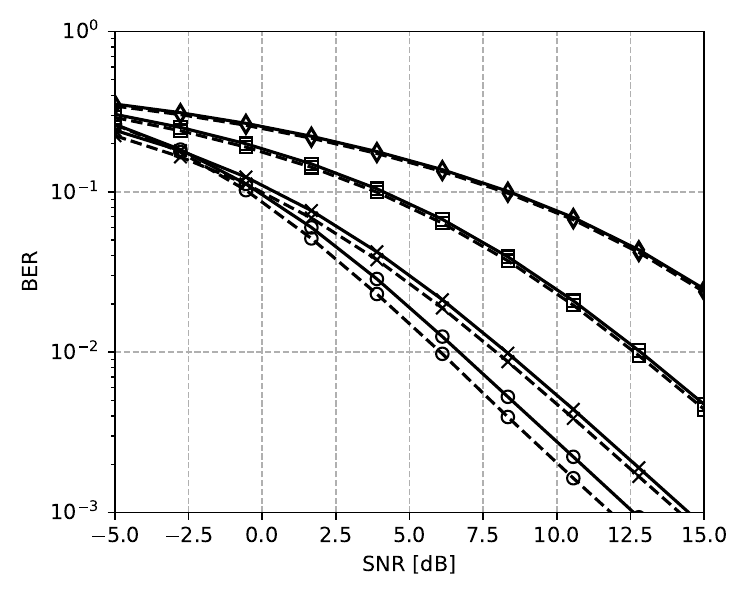}
\\[0.3cm]
\includegraphics[width=0.44\textwidth,trim=0cm 0.4cm 0cm 0.3cm, clip]{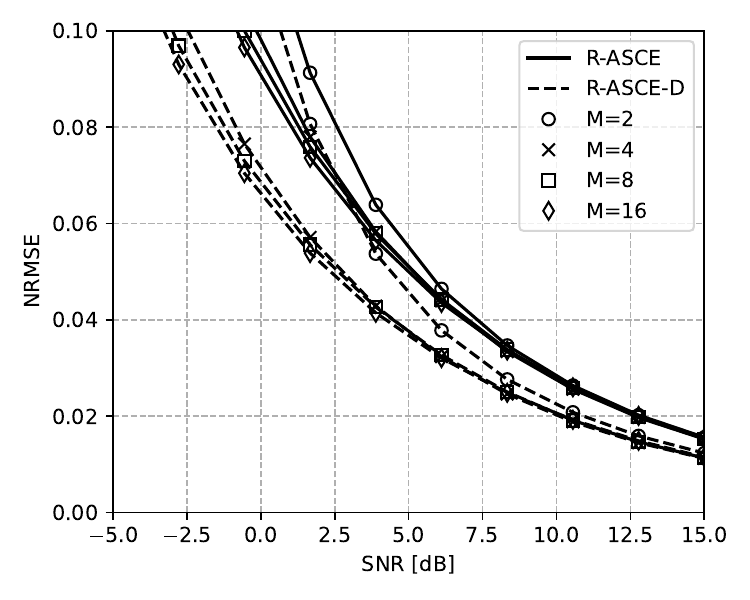}
\caption{BER (top) and NRMSE (bottom)  versus SNR when $N=1$, $L=52$, and $(P,M)=(4,2),(28,4),(36,8),(40,16)$. Other system parameters: $R=48/52$ bits/subchannel-use/tag, $\text{SIR}=-10$~dB,  $Q=3$, $K_s=8$, and $\bar K=3$.}
\label{BER_RMSE_M}
\end{figure}
Finally, Fig.~\ref{BER_RMSE_M} shows BER and NRMSE of R-ASCE and R-ASCE-D versus SNR for different values of $M$ when $L=52$ and $Q=3$. Here,  the same transmission rate of $48/52\simeq 0.92$ bits/subchannel-use/tag is maintained by scaling up the number of pilot symbols for larger $M$. BER decreases when $M$ increases since the negative effect of reducing the distance among the constellation points overcomes the positive effect of having a more accurate estimate of the subchannel response.

\subsection{Encoding across Multiple Subchannels}\label{SEC:Numerical analysis-multichannel}

\begin{figure}[!t]
\centering
\includegraphics[width=0.44\textwidth,trim=0cm 0.4cm 0cm 0.3cm, clip]{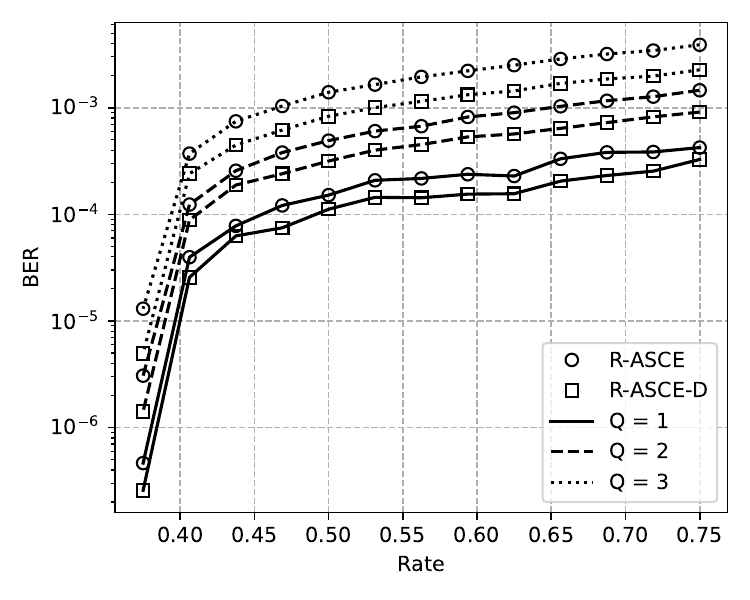}
\\[0.3cm]
\includegraphics[width=0.44\textwidth,trim=0cm 0.4cm 0cm 0.3cm, clip]{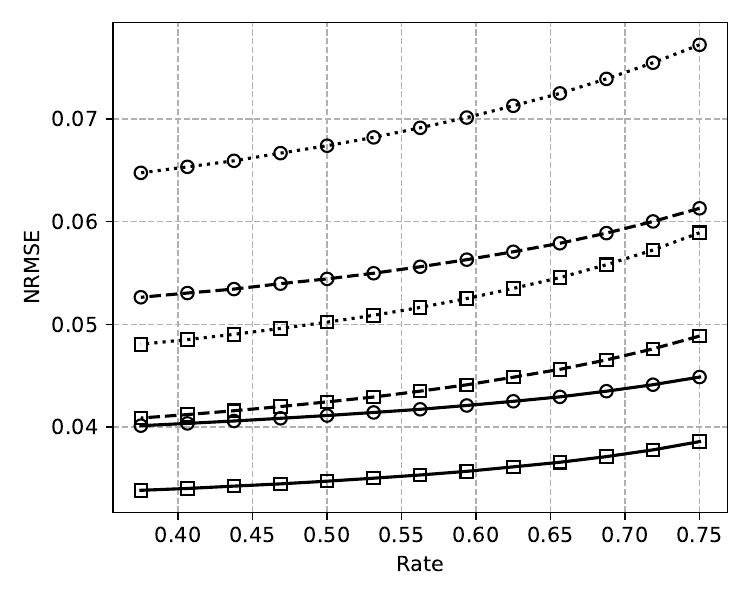}
\caption{BER (top) and NRMSE (bottom)  versus the transmission rate for $Q=1,2,3$, when $N=2$, $L=16$, $P_1=P_2=4$, and  $D_0$ is varied from $12$ (lowest rate) to $0$ (highest rate). Other system parameters: $\text{SNR}=10$~dB, $\text{SIR}=-10$~dB,  $M=2$,  $K_s=8$, and $\bar K=3$.}
\label{BER_NRMSE_vs_D_fixed_P_N=2} 
\end{figure}

\begin{figure}
\centering
\includegraphics[width=0.44\textwidth,trim=0cm 0.4cm 0cm 0.3cm, clip]{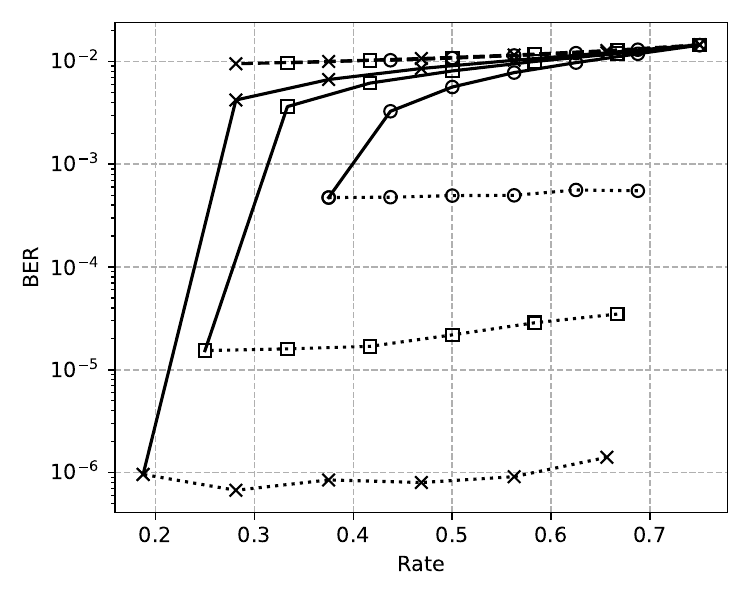}
\\[0.3cm]
\includegraphics[width=0.44\textwidth,trim=0cm 0.4cm 0cm 0.3cm, clip]{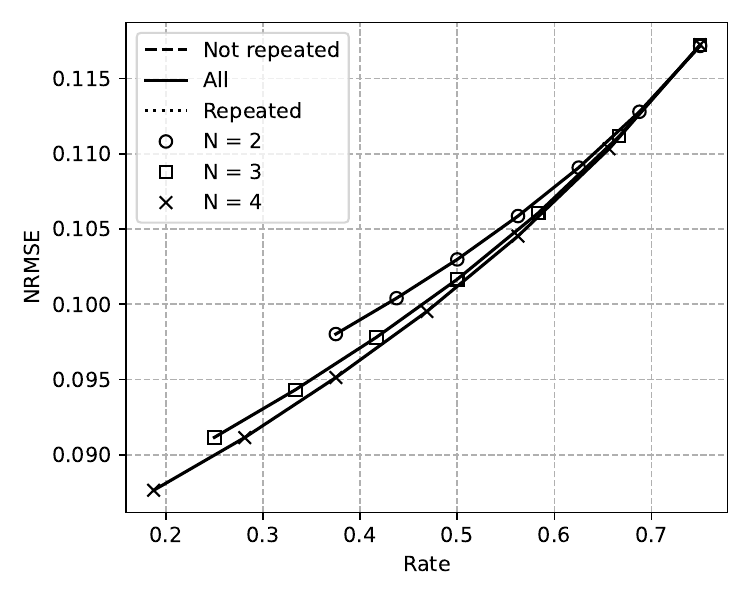}
\caption{BER (top) and NRMSE (bottom)  versus the transmission rate for $N=2,3,4$, when $L=16$, $P_1=P_2=4$, and  $D_0$ is varied from $12$ (lowest rate) to $0$ (highest rate). Other system parameters: $\text{SNR}=5$~dB, $\text{SIR}=-10$~dB,  $M=2$, $Q=2$, $K_s=8$, and $\bar K=3$.}
\label{BER_NRMSE_vs_D_fixed_P_N=234} 
\end{figure}
We now consider the joint use of multiple subchannels. We first investigate the effect of changing the transmission rate by varying the number of repeated data symbols $D_0$ from $12$ (lowest rate) to $0$ (highest rate) when $L=16$ and $P_1=P_2=4$. Fig.~\ref{BER_NRMSE_vs_D_fixed_P_N=2} reports BER and NRMSE of R-ASCE and R-ASCE-D versus the transmission rate for $Q=1,2,3$ when $\text{SNR}=10$~dB and $N=2$. In all scenarios, the conditions established by Theorem~\ref{thm:unique} are satisfied. It is again seen by inspection that R-ASCE-D outperforms R-ASCE. Also, their performance gracefully degrades as the number of tags increases, confirming their robustness against multi-tag and radar interference. Finally, BER improves as the transmission rate is reduced: this occurs not only because more repeated data symbols can enjoy diversity but also because a larger $D_0$ facilitates the estimation of the subchannel response matrices, thus resulting in a lower NRMSE. To get more insights into these latter points, for the same setup of Fig.~\ref{BER_NRMSE_vs_D_fixed_P_N=2}, Fig.~\ref{BER_NRMSE_vs_D_fixed_P_N=234} shows BER and NRMSE of R-ASCE versus the transmission rate for $N=2,3,4$ when $\text{SNR}=5$~dB and $Q=2$. In addition to the average BER (solid lines), we also report the BER of the data symbols in $\bm{D}_{0}$ that are repeated across subchannels (dotted line) and of the data symbols in $\{\bm{D}_{n}\}_{1}^{N}$ that are not repeated (dashed line). For a given transmission rate, the BER of the data symbols in $\bm{D}_{0}$ is lower than that of the data symbols in $\{\bm{D}_{n}\}_{1}^{N}$; also, both values decrease when the transmission rate decreases since a better estimate of the subchannel response matrices is obtained; finally, the average BER is in between these two values, converging to the upper and lower values for large and small transmission rates, respectively. Overall, it is seen that different tradeoffs can be obtained by changing $N$ and $D_0$; also, the upper rightmost marker point coincides with all values of $N$, as this operating point corresponds to $D_0=0$ (which is equivalent to taking $N=1$).

\begin{figure}
\centering
\includegraphics[width=0.44\textwidth,trim=0.3cm 0.2cm 0.8cm 1.3cm, clip]{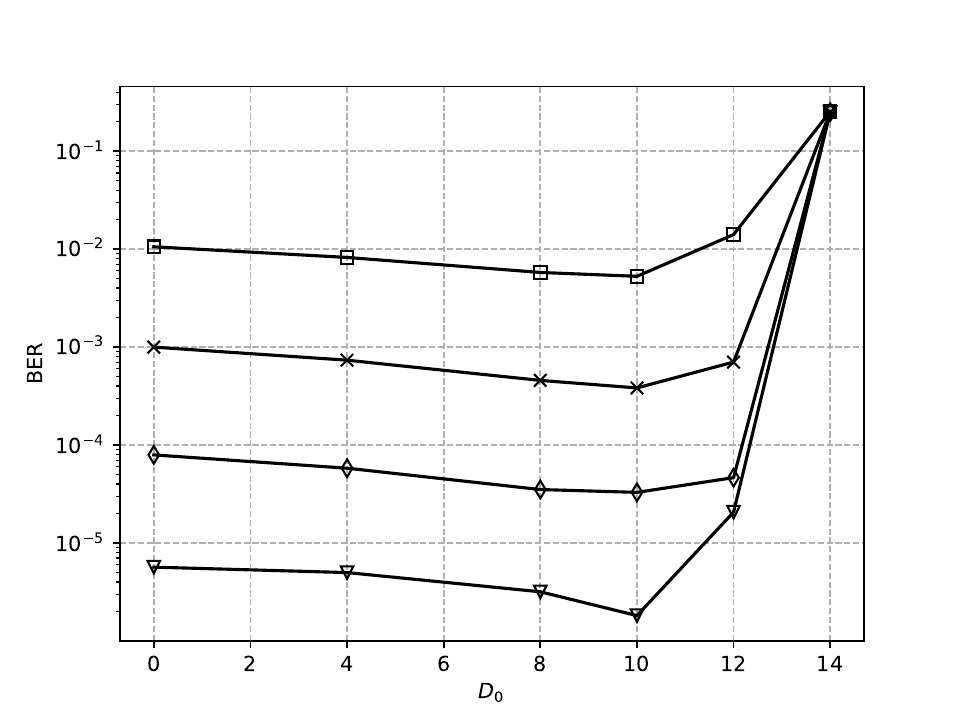}
\\[0.3cm]
\includegraphics[width=0.44\textwidth,trim=0.3cm 0.2cm 0.8cm 1.3cm, clip]{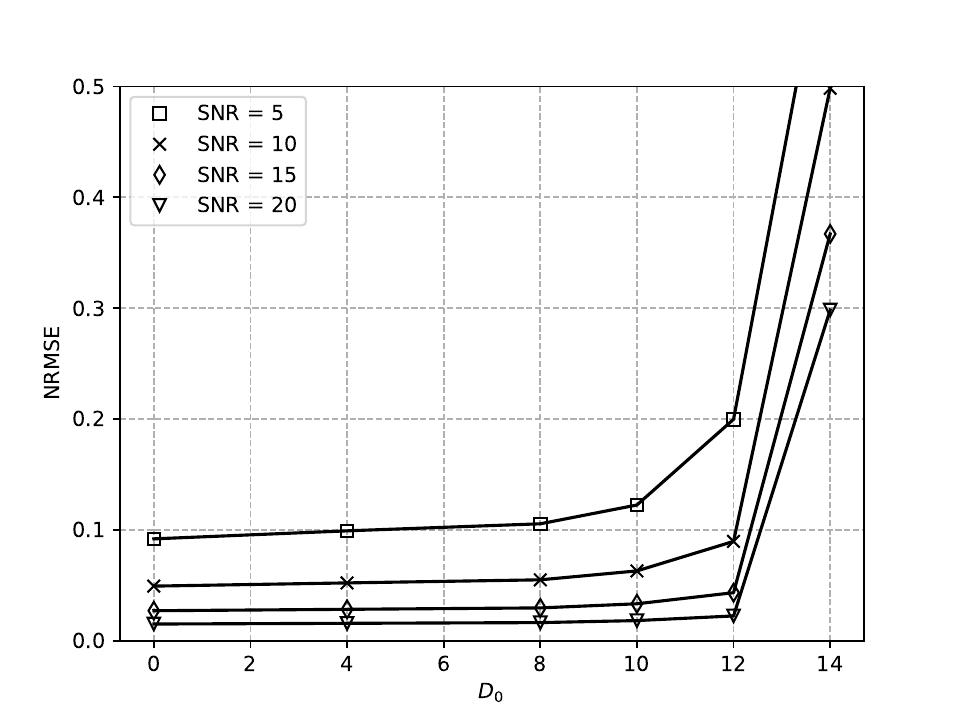}
\caption{BER (top) and NRMSE (bottom) versus even values of $D_0$ when $N=2$, $L=16$, and $P_1=P_2=(L-D_0)/2$. Other system parameters: $R=0.5$ bits/subchannel-use/tag, $\text{SIR}=-10$~dB,  $M=2$, $Q=2$, $K_s=8$, and $\bar K=3$.} 	
 \label{BER_NRMSE_vs_D0} 
\end{figure}
Next, we investigate the effect of increasing $D_0$ by reducing the pilot symbols when $L=16$ and the transmission rate is fixed to $R=0.5$ bits/subchannel-use/tag. Fig.~\ref{BER_NRMSE_vs_D0} reports BER and NRMSE of R-ASCE-D versus even values of $D_0$ for various SNR values, when $P_1=P_2=(L-D_0)/2$. For $D_0\leq 10$, we have $P_1=P_2\geq 4$ and the conditions established by Theorem~\ref{thm:unique} are satisfied: in this regime, it is beneficial to trade pilot symbols for repeated data symbols. On the other hand, when $D_0\geq 12$, we have $P_1=P_2\leq 2$, and the conditions established by Theorem~\ref{thm:unique} may not hold in some frames (this occurrence is more frequent for $D_0=14$ than for $D_0=12$): in this regime, the performance rapidly degrades.

\begin{figure}
\centering
\includegraphics[ width=0.44\textwidth, trim=0.3cm 0.2cm 0.8cm 1.3cm, clip]{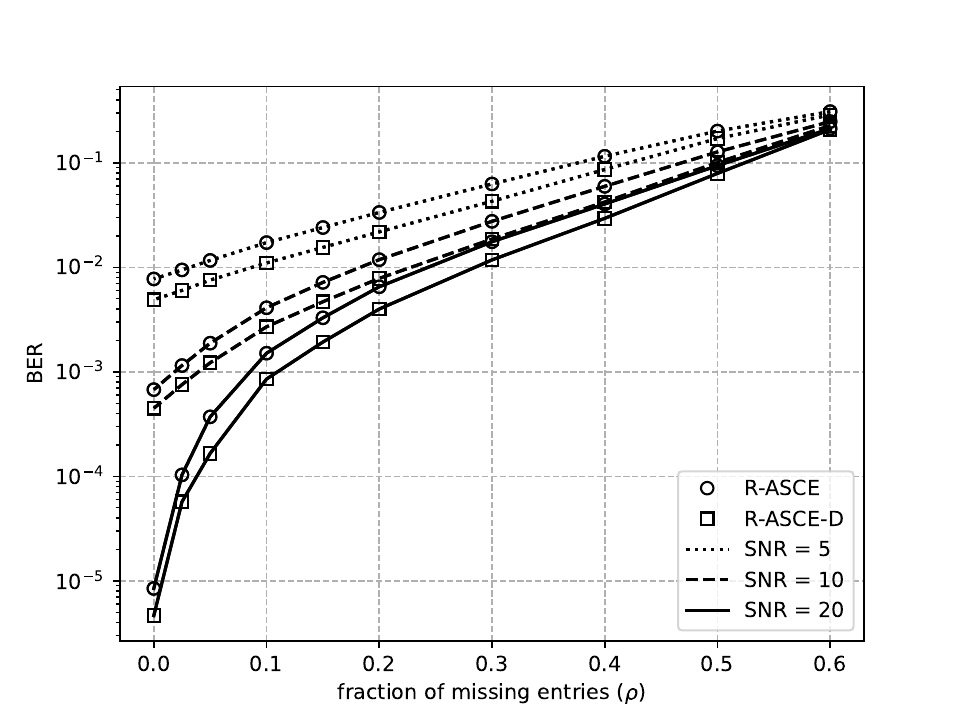}
\\[0.3cm]
\includegraphics[width=0.44\textwidth, trim=0.3cm 0.2cm 0.8cm 1.3cm, clip]{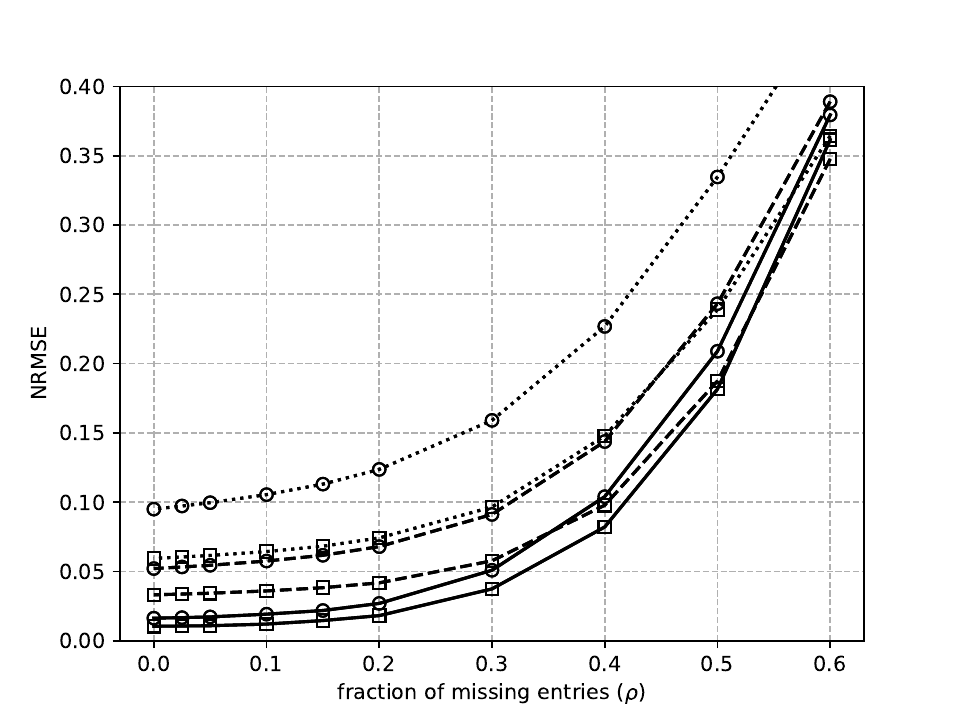}
\caption{BER (top) and NRMSE (bottom)  versus the fraction of missing entries in the measurements for $\text{SNR}=5,10,20$~dB when $N=2$, $L=16$, $P_1=P_2=4$, and $D_0=8$. Other system parameters: $R=0.5$ bits/subchannel-use/tag,  $\text{SIR}=-10$~dB, $M=2$, $Q=2$, $K_s=8$, and $\bar K=3$.}
\label{BER_NRMSE_missing_data} 
\end{figure}
Finally, Fig.~\ref{BER_NRMSE_missing_data} reports BER and NRMSE of R-ASCE and R-ASCE-D versus the fraction $\rho$ of missing entries in the measurement matrices for $\text{SNR}=5,10,20$~dB when $N=2$, $L=16$, $P_1=P_2=4$, and $D_0=8$. As expected, BER and NRMSE increase when $\rho$ is reduced.

\section{Conclusions} \label{SEC:Conclusions}
In this study, we have developed the idea of multi-tag radar-enabled backscatter communications by introducing a new encoding strategy that exploits multiple subchannels. Each tag makes use of a limited number of pilot symbols and repeats some of the data symbols over multiple subchannels to facilitate signal recovery at the reader. We have established sufficient conditions for unique data recovery, linking the number of pilot symbols to the number of active tags and highlighting the interplay among pilot and repeated data symbols. We have proposed two semi-blind decoding algorithms whose implementation is scalable with the number of tags and their payload. A thorough performance analysis is also presented, showing the diverse tradeoffs the newly proposed encoding/decoding strategies entail and eliciting the advantages over their competitors. 

We underline that these results hold true with no strong assumption on the environment the multi-tag system is required to operate in. Needless to say, further optimization would be possible should some prior information on the system operating conditions be available or {\em learnable}. The proposed architecture could indeed greatly benefit from smart sensor placement philosophies,  wherein the tags, which employ the radar clutter as carrier signals, are in close proximity to large and stationary scattering centers so as to be able to transmit long data streams with significant power on many subchannels, whereas the reader is placed in a relatively clutter-free region, possibly in line-of-sight with the tags, to limit the communication links attenuation and the ``exogenous'' radar-generated interference it is exposed to. 

\appendix[Proof of Theorem~\ref{thm:unique}]
Observe that $\bm X_n \bm A_n$ is a full rank factorization of $\bm Y_n$ by construction. If $\bm T_n \bm V_n$ is another full rank factorization of $\bm Y_n$, we must have $\bm T_n=\bm X_n \bm R_n$ and $\bm A_n=\bm R_n \bm V_n$, for some invertible matrix $\bm R_n$, $n=1,\ldots,N$ \cite[Theorem 2]{piziak1999}. The theorem is therefore proven if we show that $\bm R_n=\bm I_{Q+1}$ for every $n$. Notice that, from~\eqref{eq:Xnstructure} and~\eqref{eq:Tnstructure}, we must have
\begin{subequations}
    \begin{align}
        [\bm P_n \; \bm 1_{P_n}] &= [\bm P_n \; \bm 1_{P_n}] \bm R_n, \; \forall n, \\
        [\bm U_0 \; \bm 1_{D_0}] &= [\bm D_0 \; \bm 1_{D_0}] \bm R_n, \; \forall n,
    \end{align}\label{conds_Rn}
\end{subequations}
which in turn implies that
\begin{subequations}\label{a_conds}
    \begin{align}
    [\bm P_n \; \bm 1_{P_n}] (\bm R_n - \bm I_{Q+1})&=\bm O_{P_n, Q+1}, \; \forall n \label{a_conds-1} \\
    [\bm D_0 \; \bm 1_{D_0}] (\bm R_n - \bm R_m) &=\bm O_{D_0, Q+1}, \; \forall n \label{a_conds-2}
    \end{align} 
\end{subequations}
for any $m\in\{1,\ldots,N\}$. From~\eqref{a_conds-2}, we now have that
\begin{equation}
    \mathrm{col}(\bm R_n-\bm R_m) \subseteq \mathrm{null} \bigl( [ \bm D_0\; \bm 1_{D_0}]\bigr), \; \forall n  \label{a_cond_D}
\end{equation}
and, from~\eqref{a_conds-1}, that
\begin{align}
 \mathrm{col}(\bm R_n-\bm R_m) &= \mathrm{col}\bigl((\bm R_n- \bm I_{Q+1}) -(\bm R_m -\bm I_{Q+1}) \bigr)\notag\\
 & \subseteq \mathrm{col}(\bm R_n- \bm I_{Q+1}) + \mathrm{col}(\bm R_m- \bm I_{Q+1}) \notag \\ 
 &\subseteq \mathrm{null} \bigl( [ \bm P_n \; \bm 1_{P_n}]\bigr) + \mathrm{null} \bigl( [ \bm P_m \; \bm 1_{P_m}]\bigr), \quad \forall n. \label{a_cond_P}
\end{align}
From~\eqref{thconds-2}, Eqs.~\eqref{a_cond_D} and~\eqref{a_cond_P} can be simultaneously satisfied only if $\bm R_n=\bm R, \forall n$. At this point,~\eqref{a_conds-1} becomes
\begin{equation}
 \begin{bmatrix}
 [\bm P_1 \;\; \bm 1_{P_1}] (\bm R- \bm I_{Q+1})\\
  \vdots\\
 [\bm P_N \;\; \bm 1_{P_N}] (\bm R- \bm I_{Q+1})
 \end{bmatrix}  = [\bm P  \;\; \bm 1_{P}] (\bm R- \bm I_{Q+1}) =\bm O_{P,Q+1},
\end{equation}
which implies that $\bm R= \bm I_{Q+1}$ from condition~\eqref{thconds-1}.

\end{document}